\documentclass[smallextended]{svjour3}                     
\usepackage{amsmath}
\usepackage{amsfonts}
\usepackage{amssymb}
\usepackage{graphicx}
\usepackage{bm}
\numberwithin{equation}{section}
\journalname{Quantum Information Processing}
\begin{document}

\title{Linear and integrable nonlinear evolution of the qutrit}


\author{Krzysztof Kowalski}


\institute{Department of Theoretical Physics, University
of \L\'od\'z, ul.\ Pomorska 149/153, 90-236 \L\'od\'z,
Poland\\ \email{kowalski@uni.lodz.pl}}

\date{Received: date / Accepted: date}

\maketitle

\begin{abstract}
The nonlinear generalization of the von Neumann equation preserving convexity of 
the state space is studied in the nontrivial case of the qutrit.  This equation 
can be cast into the integrable classical Riccati system of nonlinear ordinary 
differential equations.  The solutions of such system are investigated in both 
the linear case corresponding to the standard von Neumann equation and the 
nonlinear one referring to the generalization of this equation.  The analyzed 
dynamics of the qutrit is rich and includes quasi-periodic motion, multiple 
equilibria and limit cycles.
\end{abstract}
\section{Introduction}
\label{intro}
In recent paper \cite{1} the evolution of the density matrix was studied of the form
\begin{equation}
\rho(t)=\frac{e^{t(G-{\rm i}H)}\rho_0e^{t(G+{\rm i}H)}}
{{\rm Tr}[e^{t(G-{\rm i}H)}\rho_0e^{t(G+{\rm i}H)}]},
\end{equation}
where $\rho_0$ is the $N\times N$ matrix and $H$ and $G$ are  Hermitian.  The 
local form of (1.1) is the nonlinear generalization of the von Neumann equation 
given by
\begin{equation}
\dot\rho(t)=-{\rm i}[H,\rho(t)]+\{G-{\rm Tr}[\rho(t) G],\rho(t)\},\qquad 
\rho(0)=\rho_0,
\end{equation}
where $\{\cdot,\cdot\}$ designates the anticommutator. We recall \cite{1} that the 
following nonlinear Schr\"odinger equation implied by (1.2) when $\rho(t)=|\phi(t)\rangle\langle\phi(t)|$
is the pure state
\begin{equation}
{\rm i}\frac{d}{dt}|\phi(t)\rangle = (H+{\rm i}G)|\phi(t)\rangle-{\rm i}
\langle\phi(t)|G|\phi(t)\rangle|\phi(t)\rangle,\qquad 
|\phi(0)\rangle=|\phi_0\rangle,
\end{equation}
was originally introduced by Gisin \cite{2} as a nonlinear candidate for description of 
quantum evolution of dissipative systems.  It is worthwhile to note that this equation is 
recognized as ``the only sensible candidate for a dissipative Schr\"odinger equation'' 
\cite{3}.  Another application of (1.3) written in the form
\begin{equation}
{\rm i}\frac{d}{dt}|\psi(t)\rangle = H|\psi(t)\rangle+ (1-P_{\psi(t)})U|\psi(t)\rangle, 
\end{equation}
where $P_{\psi(t)}=|\psi(t)\rangle\langle\psi(t)|$ is the projection operator and $U$ is 
an arbitrary linear or nonlinear operator, introduced by Grigorenko \cite{4} is the modelling 
the collapse of the wave function when a measurement is made on a quantum system.

An important property of the complete positive map specified by (1.1) such that
\begin{equation}
\Phi(\rho_0)=\frac{A(t)\rho_0 A^\dagger(t)}{{\rm Tr}(A(t)\rho_0 A^\dagger(t))},
\end{equation}
where $A(t)=e^{t(G-{\rm i}H)}$, is that it preserves the convex structure of the 
state space.  Namely, we have \cite{1,5,6}
\begin{equation}
\Phi[\lambda\rho_1+(1-\lambda)\rho_2]=\lambda'\Phi(\rho_1)+(1-\lambda')\Phi(\rho_2),
\end{equation}
where $\rho_1$ and $\rho_2$ are density matrices, and 
\begin{equation}
\lambda'=\frac{{\rm Tr}(A\rho_1A^\dagger)}{{\rm Tr}\{A[\lambda\rho_1+(1-\lambda)\rho_2]
A^\dagger\}}\lambda
\end{equation}
satisfies $\lambda'\in[0,1]$.  Another relevant feature of the dynamics 
described by Eq.\ (1.1) is that it maps pure states into pure ones.  We remark that
non-unitary evolution in general takes pure states to mixed states. Finally, a desired property of 
the investigated nonlinear generalization of the von Neumann equation is that it does not allow 
superluminal messages \cite{7}.  We recall that the superluminal signalling is one of 
the crucial arguments against nonlinear generalizations of quantum mechanics such as for example
the Schr\"odinger-Newton equation \cite{8}.

We point out that (1.1) is actually the linearization transformation for (1.2).  Indeed, the 
numerator of (1.1) satisfies the linear equation
\begin{equation}
\dot{\tilde\rho}(t)=-{\rm i}[H,{\tilde\rho}(t)]+\{G,{\tilde\rho}(t)\},\qquad 
\rho(0)=\rho_0,
\end{equation}
where ${\tilde\rho}(t)=e^{t(G-{\rm i}H)}\rho_0e^{t(G+{\rm i}H)}$, so the solution 
of the nonlinear von Neumann equation (1.2) can be cast into the solution of the 
linear equation (1.8).  Notice that (1.8) with $G$ replaced by $-\frac{1}{2}G$ 
coincides with the Lamb equation \cite{9} utilized in the theory of the 
optical maser.

The quantum dynamics described by the nonlinear equation (1.2) was illustrated in 
\cite{1} by the example of the qubit. To be more specific, we set
\begin{equation}
\rho_0=\frac{1}{2}(1+{\bm \xi}_0\cdot{\bm\sigma}),
\end{equation}
where $\sigma_i$, $i=1,\,2,\,3$ are the Pauli matrices, the dot designates the scalar product, 
the Bloch vector ${\bm\xi}_0\in{\mathbb R}^3$ satisfies 
$|{\bm\xi}_0|\leqslant1$, and
\begin{equation}
H={\bm a}\cdot{\bm\sigma},\qquad  G={\bm b}\mbox{\boldmath$
\scriptstyle{\cdot}$}{\bm\sigma}.
\end{equation}
On substituting (1.9) and (1.10) into (1.1) we arrive at the following form of the 
density matrix $\rho(t)$
\begin{equation}
\rho(t)=\frac{1}{2}(1+{\bm \xi}(t)\cdot{\bm\sigma}),
\end{equation}
where ${\bm \xi}(t)$ is an explicit  function of $t$ and ${\bm \xi}_0$ (see 
\cite{1}).  On the other hand, inserting (1.11) into the nonlinear von Neumann 
equation (1.2) we find that ${\bm \xi}(t)$ is the solution of the following 
nonlinear system of ordinary differential equations
\begin{equation}
\dot{\bm\xi}=2{\bm b}+2{\bm a}\times{\bm\xi}-2({\bm b}\cdot
{\bm\xi}){\bm\xi},\qquad {\bm\xi}(0)={\bm\xi}_0. 
\end{equation}
Thus, it turns out that the nonlinear quantum evolution equation (1.2) can be 
reduced to the nonlinear classical system (1.12).  The system (1.12) with quadratic 
nonlinearities is called the Riccati system.  We remark that as with (1.2) the system 
(1.12) is quasi-linear.  The linearization transformation is of the form \cite{1}
\begin{equation}
{\bm \xi}(t)=\frac{{\bm\eta}(t)}{\varphi(t)},
\end{equation}
where
\begin{equation}
\begin{split}
\dot\varphi&=2{\bm b}\cdot{\bm \eta},\\
\dot{\bm \eta}&=2\varphi{\bm b}+2{\bm a}\times{\bm \eta},\qquad 
\varphi(0)=1,\,\,{\bm \eta}(0)={\bm \xi}_0.
\end{split}
\end{equation}
The solutions to (1.12) were analyzed in a great detail in \cite{1}.  In 
particular an interesting property of the nonlinear dynamics of the system (1.12) 
was found --- the global asymptotic stability of stationary (equilibrium) 
solutions corresponding to evolution of the qubit from mixed states to pure ones.

An example of a physical application of the discussed approach is the 
relativistic quantum spin one-half particle in  electromagnetic field analyzed in Ref.\ 1.  
More precisely, the evolution is introduced therein such that the $\rho(t)$ and $A(t)$ in 
(1.5) are connected with the Bargmann-Michel-Telegdi equations \cite{10} 
describing relativistic motion of a particle with a magnetic moment in the 
external electromagnetic field.  The vectors ${\bm a}$ and ${\bm b}$ from (1.10) 
are identified with the external magnetic and electric field, respectively.  
Another interesting physical example is the quantum dynamics with the $su(1,1)$ 
Hamiltonian given by (1.10) with time-dependent ${\bm a}$ and ${\bm b}$ 
introduced in Ref.\ 11, regarded as a Rabi problem with a complex transverse 
magnetic field.  Finally, utilizing a nonlinear generalization of the 
Gorini-Kossakowski-Sudarshan-Lindblad equation in the case of the qubit, based on the 
nonlinear von Neumann equation (1.2) as a point of departure, the extension of the 
celebrated Jaynes-Cummings model was introduced in Ref.\ 12 describing the interaction 
of a two-level atom with a single mode of the electromagnetic field.

In this work following the approach taken up in the case of the qubit we study 
the much more complicated case of the qutrit.  The paper is organized as follows.
In Sect.\  2 we introduce the nonlinear system that is the counterpart of (1.12) 
in the case with the qutrit.  Section 3 is devoted to the linear case referring 
to the standard von Neumann equation.  Section 4 deals with the nonlinear system.
All the necessary identities corresponding to the $su(3)$ algebra are collected 
in Appendix.
\section{Nonlinear evolution of a qutrit}
\label{sec:2}
Consider now the special case of a qutrit.  The well known generalized Bloch 
sphere representation for a qutrit is specified by the density matrix
\begin{equation}
\rho=\frac{1}{3}(1+\sqrt{3}{\bm\xi}\cdot{\bm\lambda}),
\end{equation}
where $\lambda_i$, $i=1,\,2,\ldots,\,8$, are the Gell-Mann matrices (see Appendix), 
and the state space $\Omega$ is given by \cite{13}
\begin{equation}
\Omega=\{{\bm\xi}\in{\mathbb R}^8:\,  
3{\bm\xi}^2-2{\bm\xi}\cdot({\bm\xi}*{\bm\xi})\leqslant1,\quad 
{\bm\xi}^2\leqslant1\},
\end{equation}
where ${\bm a}*{\bm b}$ is the symmetric star product of vectors ${\bm a}$, 
${\bm b}\in{\mathbb R}^8$ described in Appendix.  In opposition to the qubit 
states the boundary $\partial\Omega$ of the space of states $\Omega$ specified by
\begin{equation}
\partial\Omega=\{{\bm\xi}\in{\mathbb R}^8:\, 3{\bm\xi}^2-2{\bm\xi}\cdot({\bm\xi}*{\bm\xi})=1,\quad 
{\bm\xi}^2\leqslant1\},
\end{equation} 
contains both the pure states characterized by
\begin{equation}
\partial\Omega_p=\{{\bm\xi}\in{\mathbb R}^8:\,{\bm\xi}^2=1,\quad{\bm\xi}*{\bm\xi}={\bm\xi}\},
\end{equation}
and mixed states:
\begin{equation}
\partial\Omega_m=\{{\bm\xi}\in{\mathbb R}^8:\,  
3{\bm\xi}^2-2{\bm\xi}\cdot({\bm\xi}*{\bm\xi})=1,\quad 
{\bm\xi}^2<1\},
\end{equation}
The state space $\Omega$ is five-dimensional and its boundary $\partial\Omega$ is 
four-dimensional.  Taking into account that $\partial\Omega_p$ is the coset space 
$SU(3)/U(2)$ one can find the following useful parametrization of $\partial\Omega_p$
by angles $\alpha$, $\beta$, $\gamma$ and $\delta$
\begin{equation}
\begin{split}
{\bm\xi}=\sqrt{3}(&\sin\alpha\sin\beta\cos\alpha\cos(\delta-\gamma),\sin\alpha\sin\beta\cos\alpha
\sin(\delta-\gamma),\\
&\frac{1}{2}(\cos^2\alpha\sin^2\beta-\sin^2\alpha),\cos^2\alpha\cos\beta\sin\beta\cos\gamma,\\
&-\cos^2\alpha\cos\beta\sin\beta\sin\gamma,\cos\beta\cos\alpha\sin\alpha\cos\delta,\\
&-\cos\beta\cos\alpha\sin\alpha\sin\delta,\frac{1}{2\sqrt{3}}(\cos^2\alpha\sin^2\beta+\sin^2\alpha-
2\cos^2\alpha\cos^2\beta)).
\end{split}
\end{equation}
We remark that there exist in the literature the alternative parametrizations of $\partial\Omega_p$
(see for example \cite{14}).

Now proceeding analogously as with the qubit states we set
\begin{equation}
\rho_0=\frac{1}{3}(1+\sqrt{3}{\bm\xi}_0\cdot{\bm\lambda}),\qquad 
\rho(t)=\frac{1}{3}(1+\sqrt{3}{\bm\xi}(t)\cdot{\bm\lambda}),
\end{equation}
and
\begin{equation}
H={\bm a}\cdot{\bm\lambda}, \qquad G={\bm b}\cdot{\bm\lambda}.
\end{equation}
On substituting (2.7) and (2.8) into (1.2) and using the identities (A.30), (A.32) and (A.33) we arrive 
at the following Riccati system of nonlinear ordinary differential equations
\begin{equation}
\dot{\bm\xi}=\frac{2}{\sqrt{3}}{\bm b}+\frac{2}{\sqrt{3}}{\bm 
a}\wedge{\bm\xi}+\frac{2}{\sqrt{3}}{\bm b}*{\bm \xi}-\frac{4}{\sqrt{3}}
({\bm b}\cdot{\bm\xi}){\bm\xi},\qquad {\bm\xi}(0)={\bm\xi}_0,
\end{equation}
where ${\bm a}\wedge{\bm b}$ is the antisymmetric product of vectors ${\bm a}$ 
and ${\bm b}$ (see Appendix).  It thus appears that the nonlinear quantum evolution 
equation (1.2) with $\rho(t)$, $H$ and $G$ given by (2.7) and (2.8), 
respectively, reduces to the nonlinear classical system (2.9).
\section{Linear evolution of a qutrit: periodic and quasiperiodic solutions}
\label{sec:3}
We now restrict to the case ${\bm b}=0$.  The system (2.9) reduces then to the
linear one corresponding to the linear von Neumann equation.  Namely, we have
\begin{equation}
\dot{\bm\xi}=\frac{2}{\sqrt{3}}{\bm a}\wedge{\bm\xi},\qquad {\bm\xi}(0)={\bm\xi}_0
\end{equation}
In opposition to the evolution of the qubit discussed in \cite{1} even in such a 
linear case the problem of finding the solution to (3.1) by means of the 
transformation (1.1) is in general complex and leads to cumbersome formulas.  To 
be more specific consider first the problem of calculating the exponential 
$e^{{\bm a}\cdot{\bm\lambda}}$.  Referring to (1.1) and (2.1) we point out that in view 
of (A.19) and (A.24) an arbitrary power of ${\bm a}\cdot{\bm\lambda}$ has the expansion of the
form
\begin{equation}
({\bm a}\cdot{\bm\lambda})^n=c_n+d_n{\bm a}\cdot{\bm\lambda}+e_n({\bm a}*{\bm a})\cdot
{\bm \lambda},\qquad n=0,\,1,\,2,\,\ldots,
\end{equation}
where $c_n$, $d_n$ and $e_n$ are the scalar coefficients.  On using (A.19), (A.24) and (A.30) we arrive 
at the following system of recurrence equation
\begin{align}
c_{n+1}&=\frac{2}{3}{\bm a}^2d_n+\frac{2}{3}{\bm a}\cdot({\bm a}*{\bm a})e_n,\nonumber\\
d_{n+1}&=\frac{1}{\sqrt{3}}{\bm a}^2e_n+c_n,\\
e_{n+1}&=\frac{1}{\sqrt{3}}d_n,\nonumber
\end{align}
subject to the initial data $c_0=1$, $c_1=0$, $d_0=0$, $d_1=1$, $e_0=0$, $e_1=0$. 
From (3.3) we get the recursive relation
\begin{equation}
d_{n+3}-{\bm a}^2d_{n+1}-\frac{2}{3\sqrt{3}}{\bm a}\cdot({\bm a}*{\bm a})d_n=0.
\end{equation}
The characteristic equation corresponding to (3.4) is of the form
\begin{equation}
x^3-{\bm a}^2x-\frac{2}{3\sqrt{3}}{\bm a}\cdot({\bm a}*{\bm a})=0.
\end{equation} 
The discriminant of the cubic equation (3.5) is
\begin{equation}
Q=\frac{1}{27}\{[{\bm a}\cdot({\bm a}*{\bm a})]^2-|{\bm a}|^6\},
\end{equation}
where $|{\bm a}|$ designates the norm of the vector ${\bm a}$.  The discriminant (3.6) fulfills $Q\leqslant0$.
The simplest case $Q=0$ refers to the condition ${\bm a}*{\bm a}=\pm|{\bm a}|{\bm a}$ and will be 
discussed later.  For $Q<0$ we have the trigonometric solution \cite{15} to (3.5) 
such that
\begin{equation}
\begin{split}
x_1&=\frac{2}{\sqrt{3}}|{\bm a}|\cos\frac{\alpha}{3},\\
x_{2,3}&=-\frac{2}{\sqrt{3}}|{\bm a}|\cos\left(\frac{\alpha}{3}\pm\frac{\pi}{3}\right),
\end{split}
\end{equation}
where
\begin{equation}
\cos\alpha=\frac{{\bm a}\cdot({\bm a}*{\bm a})}{|{\bm a}|^3}.
\end{equation}
Using (3.3), (3.4) and (3.7) we find after some calculation
\begin{align}
c_n=&\frac{1}{12}\frac{4\cos^2\frac{\alpha}{3}-1}{\cos\left(\frac{\alpha}{3}+\frac{\pi}{6}\right)
\cos\left(\frac{\alpha}{3}-\frac{\pi}{6}\right)}\left(\frac{2}{\sqrt{3}}|{\bm a}|\cos\frac{\alpha}{3}\right)^n\nonumber\\
&{}+\frac{1}{12}\frac{-4\sin^2\left(\frac{\alpha}{3}-\frac{\pi}{6}\right)+1}{\cos\left(\frac{\alpha}{3}+\frac{\pi}{6}\right)
\sin\frac{\alpha}{3}}\left[-\frac{2}{\sqrt{3}}|{\bm a}|\cos\left(\frac{\alpha}{3}+\frac{\pi}{3}\right)\right]^n\\
&{}+\frac{1}{12}\frac{4\sin^2\left(\frac{\alpha}{3}+\frac{\pi}{6}\right)-1}{\cos\left(\frac{\alpha}{3}-\frac{\pi}{6}\right)
\sin\frac{\alpha}{3}}\left[-\frac{2}{\sqrt{3}}|{\bm 
a}|\cos\left(\frac{\alpha}{3}-\frac{\pi}{3}\right)\right]^n\nonumber,
\end{align}
\begin{align}
d_n=&\frac{1}{2\sqrt{3}}\frac{\cos\frac{\alpha}{3}}{|{\bm a}|\cos\left(\frac{\alpha}{3}+\frac{\pi}{6}\right)
\cos\left(\frac{\alpha}{3}-\frac{\pi}{6}\right)}\left(\frac{2}{\sqrt{3}}|{\bm a}|\cos\frac{\alpha}{3}\right)^n\nonumber\\
&{}-\frac{1}{2\sqrt{3}}\frac{\sin\left(\frac{\alpha}{3}-\frac{\pi}{6}\right)}{|{\bm a}|\cos\left(\frac{\alpha}{3}+\frac{\pi}{6}\right)
\sin\frac{\alpha}{3}}\left[-\frac{2}{\sqrt{3}}|{\bm a}|\cos\left(\frac{\alpha}{3}+\frac{\pi}{3}\right)\right]^n\\
&{}-\frac{1}{2\sqrt{3}}\frac{\sin\left(\frac{\alpha}{3}+\frac{\pi}{6}\right)}{|{\bm a}|\cos\left(\frac{\alpha}{3}-\frac{\pi}{6}\right)
\sin\frac{\alpha}{3}}\left[-\frac{2}{\sqrt{3}}|{\bm 
a}|\cos\left(\frac{\alpha}{3}-\frac{\pi}{3}\right)\right]^n\nonumber,
\end{align}
\begin{align}
e_n=&\frac{1}{4\sqrt{3}}\frac{1}{{\bm a}^2\cos\left(\frac{\alpha}{3}+\frac{\pi}{6}\right)
\cos\left(\frac{\alpha}{3}-\frac{\pi}{6}\right)}\left(\frac{2}{\sqrt{3}}|{\bm a}|\cos\frac{\alpha}{3}\right)^n\nonumber\\
&{}-\frac{1}{4\sqrt{3}}\frac{1}{{\bm a}^2\cos\left(\frac{\alpha}{3}+\frac{\pi}{6}\right)
\sin\frac{\alpha}{3}}\left[-\frac{2}{\sqrt{3}}|{\bm a}|\cos\left(\frac{\alpha}{3}+\frac{\pi}{3}\right)\right]^n\\
&{}+\frac{1}{4\sqrt{3}}\frac{1}{{\bm a}^2\cos\left(\frac{\alpha}{3}-\frac{\pi}{6}\right)
\sin\frac{\alpha}{3}}\left[-\frac{2}{\sqrt{3}}|{\bm 
a}|\cos\left(\frac{\alpha}{3}-\frac{\pi}{3}\right)\right]^n\nonumber.
\end{align}
An immediate consequence of (3.9), (3.10) and (3.11) is the following formula for 
the exponential $e^{\tau{\bm a}\cdot{\bm \lambda}}$
\begin{align}
e^{\tau{\bm a}\cdot{\bm \lambda}}=&
\frac{1}{12}\frac{4\cos^2\frac{\alpha}{3}-1}{\cos\left(\frac{\alpha}{3}+\frac{\pi}{6}\right)
\cos\left(\frac{\alpha}{3}-\frac{\pi}{6}\right)}\exp\left(\frac{2\tau}{\sqrt{3}}|{\bm a}|\cos\frac{\alpha}{3}\right)\nonumber\\
&{}+\frac{1}{12}\frac{-4\sin^2\left(\frac{\alpha}{3}-\frac{\pi}{6}\right)+1}{\cos\left(\frac{\alpha}{3}+\frac{\pi}{6}\right)
\sin\frac{\alpha}{3}}\exp\left[-\frac{2\tau}{\sqrt{3}}|{\bm a}|\cos\left(\frac{\alpha}{3}+\frac{\pi}{3}\right)\right]\nonumber\\
&{}+\frac{1}{12}\frac{4\sin^2\left(\frac{\alpha}{3}+\frac{\pi}{6}\right)-1}{\cos\left(\frac{\alpha}{3}-\frac{\pi}{6}\right)
\sin\frac{\alpha}{3}}\exp\left[-\frac{2\tau}{\sqrt{3}}|{\bm a}|\cos\left(\frac{\alpha}{3}-\frac{\pi}{3}\right)\right]\nonumber\\
&{}+\left(\frac{1}{2\sqrt{3}}\frac{\cos\frac{\alpha}{3}}{|{\bm a}|\cos\left(\frac{\alpha}{3}+\frac{\pi}{6}\right)
\cos\left(\frac{\alpha}{3}-\frac{\pi}{6}\right)}\exp\left(\frac{2\tau}{\sqrt{3}}|{\bm a}|\cos\frac{\alpha}{3}\right)\right.\nonumber\\
&{}-\frac{1}{2\sqrt{3}}\frac{\sin\left(\frac{\alpha}{3}-\frac{\pi}{6}\right)}{|{\bm a}|\cos\left(\frac{\alpha}{3}+\frac{\pi}{6}\right)
\sin\frac{\alpha}{3}}\exp\left[-\frac{2\tau}{\sqrt{3}}|{\bm a}|\cos\left(\frac{\alpha}{3}+\frac{\pi}{3}\right)\right]\nonumber\\
&{}\left.-\frac{1}{2\sqrt{3}}\frac{\sin\left(\frac{\alpha}{3}+\frac{\pi}{6}\right)}{|{\bm a}|\cos\left(\frac{\alpha}{3}-\frac{\pi}{6}\right)
\sin\frac{\alpha}{3}}\exp\left[-\frac{2\tau}{\sqrt{3}}|{\bm a}|\cos\left(\frac{\alpha}{3}-\frac{\pi}{3}\right)\right]\right){\bm a}\cdot{\bm\lambda}\nonumber\\
&{}+\left(\frac{1}{4\sqrt{3}}\frac{1}{{\bm a}^2\cos\left(\frac{\alpha}{3}+\frac{\pi}{6}\right)
\cos\left(\frac{\alpha}{3}-\frac{\pi}{6}\right)}\exp\left(\frac{2}{\sqrt{3}}|{\bm a}|\cos\frac{\alpha}{3}\right)\right.\nonumber\\
&{}-\frac{1}{4\sqrt{3}}\frac{1}{{\bm a}^2\cos\left(\frac{\alpha}{3}+\frac{\pi}{6}\right)
\sin\frac{\alpha}{3}}\exp\left[-\frac{2\tau}{\sqrt{3}}|{\bm a}|\cos\left(\frac{\alpha}{3}+\frac{\pi}{3}\right)\right]\nonumber\\
&{}\left.+\frac{1}{4\sqrt{3}}\frac{1}{{\bm a}^2\cos\left(\frac{\alpha}{3}-\frac{\pi}{6}\right)
\sin\frac{\alpha}{3}}\exp\left[-\frac{2\tau}{\sqrt{3}}|{\bm a}|\cos\left(\frac{\alpha}{3}-\frac{\pi}{3}\right)\right]\right)
({\bm a}*{\bm a})\cdot{\bm\lambda}.
\end{align}
As a matter of fact, setting $\tau=-{\rm i}t$ in (3.12) and making use of (1.1), 
(2.7) and (2.8) one can obtain the solution to (3.1).  Nevertheless, 
as mentioned earlier the formulas are cumbersome and we decided not to present 
them herein.  The numerical integration of (3.1) shows that the typical trajectory is
the quasiperiodic one such as that depicted in Fig.\ 1.  We remark that quasiperiodic 
trajectories were absent in the case with the evolution of the qubit \cite{1}.  The remaining
cases include periodic (see Fig.\ 2) and stationary solutions.  In the next sections we introduce explicit 
solutions to (3.1) in the special cases with ${\bm a}*{\bm a}=\pm|{\bm a}|{\bm a}$, 
${\bm a}\cdot({\bm a}*{\bm a})=0$, and the diagonal generator ${\bm a}\cdot{\bm\lambda}$.
\begin{figure*}
\centering
\begin{tabular}{c@{}c}
\includegraphics[width =.49\textwidth]{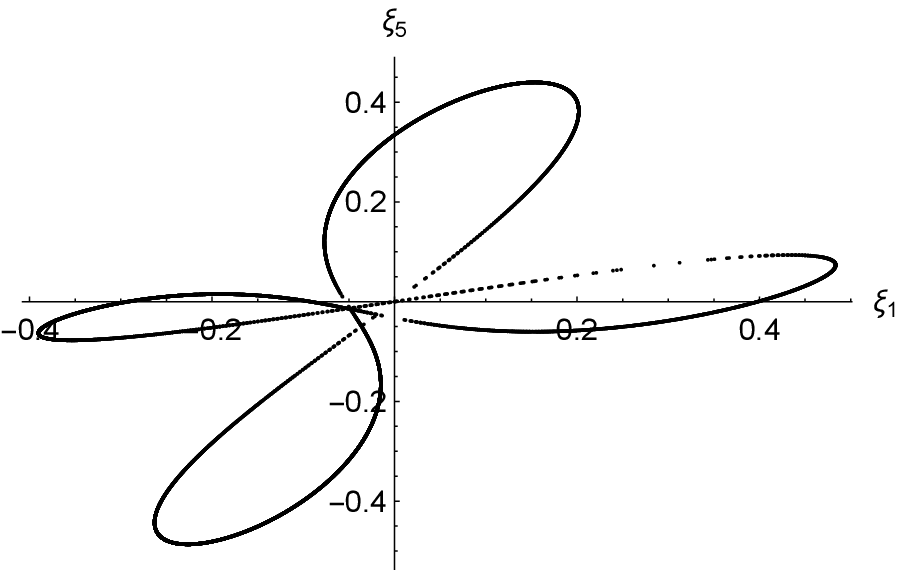}&
\includegraphics[width =.49\textwidth]{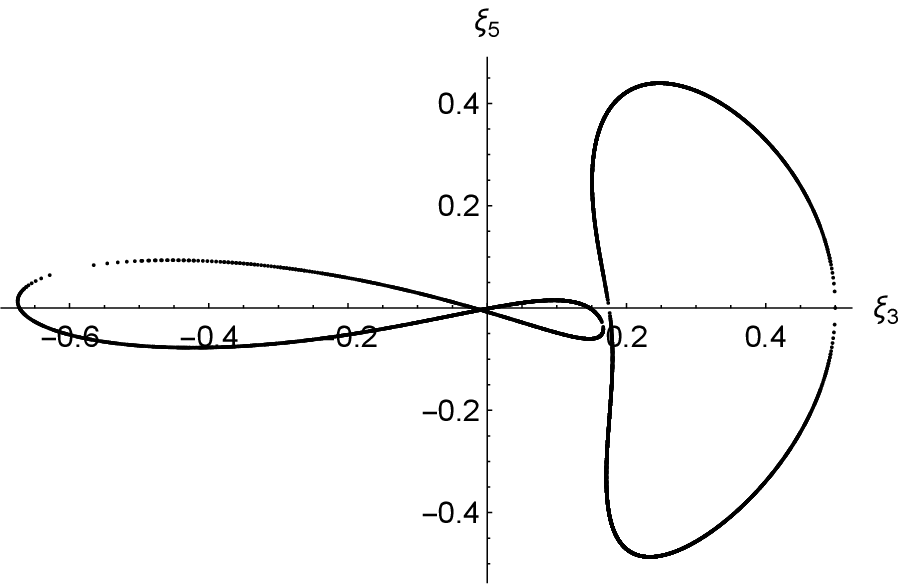}
\end{tabular}
\caption{The quasiperiodic trajectory of the system (3.1) with ${\bm a}=(0,1,0,-1,0.3,0,1,0)$ and 
${\bm\xi}_0=(0,0,\frac{1}{2},0,0,0,0,\frac{1}{2})$.  Left: the projection of the Poincar\'e section 
of the trajectory onto the $(\xi_1,\xi_5)$ plane. Right: the projection of the Poincar\'e section 
onto the $(\xi_3,\xi_5)$ plane.  The Poincar\'e section is defined by the 
hyperplane ${\bm n}\cdot({\bm\xi}-{\bm\xi}^{(0)})=0$, where the normal vector ${\bm n}=(0,1,0,0,0,0,0,0)$ 
and ${\bm\xi}^{(0)}=(0,0,0,0,0,0,0,0)$.}
\end{figure*}

We now return to (3.1).  Using (A.26) we find ${\bm\xi}^2={\rm const}$.  Further, 
taking into account (A.18), (A.26) and (A.27) we can demonstrate (see also \cite{16}) that 
${\bm\xi}\cdot({\bm\xi}*{\bm\xi})$ is the first integral of (3.1) as well. In view 
of (2.2), (2.4) and (2.5) we conclude that under the unitary linear evolution 
given by the generator ${\bm a}\cdot{\bm\lambda}$, the pure states remain pure 
ones --- this was shown in the case of the general nonlinear evolution described 
globally by Eq.\ (1.1).  Furthermore, it turns out that mixed states are also 
stable under the linear unitary evolution including the case of the mixed states 
on the boundary (2.5) $\partial\Omega_m$.

Consider now the stationary solutions to (3.1).  By virtue of (A.20) the most general 
form of these solutions is
\begin{equation}
\bar{\bm\xi}=\mu {\bm a}+\nu{\bm a}*{\bm a},
\end{equation}
where $\mu$ and $\nu$ are constants.  From (2.2) it follows that the 
states $\bar{\bm\xi}$ correspond to the following region in the $(\mu,\nu)$ plane
\begin{subequations}
\begin{align}
&{\bm a}^2\mu^2+|{\bm a}|^4\nu^2+2{\bm a}\cdot({\bm a}*{\bm a})\mu\nu\leqslant 
1,\\
&-2{\bm a}\cdot({\bm a}*{\bm a})\mu^3+2\{|{\bm a}|^6-2[{\bm a}\cdot({\bm a}*{\bm a})]^2\}
\nu^3-6|{\bm a}|^4\mu^2\nu+3{\bm a}^2\mu^2+3|{\bm a}|^4\nu^2\nonumber\\
&{}+6{\bm a}\cdot({\bm a}*{\bm a})(1-{\bm a}^2)\mu\nu\leqslant 1.
\end{align}
\end{subequations}
The pure states (2.4) lying in the boundary $\partial\Omega_p$ are given by (3.13), 
where $\mu$ and $\nu$ are the solution of the system
\begin{subequations}
\begin{align}
&{\bm a}^2\mu^2+|{\bm a}|^4\nu^2+2{\bm a}\cdot({\bm a}*{\bm a})\mu\nu=1,\\
&{\bm a}\cdot({\bm a}*{\bm a})\mu^3+\{2[{\bm a}\cdot({\bm a}*{\bm a})]^2-|{\bm a}|^6\}
\nu^3+3|{\bm a}|^4\mu^2\nu+3{\bm a}^2{\bm a}\cdot({\bm a}*{\bm a})\mu\nu=1,
\end{align}
\end{subequations}
where the second equation of (3.15) is a consequence of the relation $\bar{\bm\xi}\cdot
(\bar{\bm\xi}*\bar{\bm\xi})=1$.  Finally, the mixed states (2.5) on the boundary 
$\partial\Omega_m$ are specified by
\begin{subequations}
\begin{align}
&{\bm a}^2\mu^2+|{\bm a}|^4\nu^2+2{\bm a}\cdot({\bm a}*{\bm a})\mu\nu<1,\\
&-2{\bm a}\cdot({\bm a}*{\bm a})\mu^3+2\{|{\bm a}|^6-2[{\bm a}\cdot({\bm a}*{\bm a})]^2\}
\nu^3-6|{\bm a}|^4\mu^2\nu+3{\bm a}^2\mu^2+3|{\bm a}|^4\nu^2\nonumber\\
&{}+6{\bm a}\cdot({\bm a}*{\bm a})(1-{\bm a}^2)\mu\nu=1.
\end{align}
\end{subequations}
The concrete examples of the stationary solutions to the system (3.1) are 
discussed in the following sections.
\subsection{The case of ${\bm a}*{\bm a}=|{\bm a}|{\bm a}$}
\label{sec:3.1}
We first recall that the condition ${\bm a}*{\bm a}=|{\bm a}|{\bm a}$ refers to 
$Q=0$ in (3.6).  Taking into account (3.2) we find that in this case
\begin{equation}
({\bm a}\cdot{\bm\lambda})^n=c_n+d_n{\bm a}\cdot{\bm\lambda},\qquad 
n=0,\,1,\,2,\,\ldots. 
\end{equation}
Hence, proceeding as with (3.2) and using (A.31) we 
obtain the system of recurrence equations satisfied by $c_n$ and $d_n$.  Solving 
the second order recurrence satisfied by $d_n$ we get
\begin{equation}
e^{\tau{\bm a}\cdot{\bm\lambda}}=\frac{1}{3}e^{\frac{2\sqrt{3}}{3}\tau|{\bm a}|}+
\frac{2}{3}e^{-\frac{\sqrt{3}}{3}\tau|{\bm a}|}+\frac{1}{\sqrt{3}|{\bm a}|}
\left(e^{\frac{2\sqrt{3}}{3}\tau|{\bm a}|}-e^{-\frac{\sqrt{3}}{3}\tau|{\bm 
a}|}\right){\bm a}\cdot{\bm\lambda}.
\end{equation}
On putting $\tau=-{\rm i}t$ and making use of (1.1), (2.7) and (2.8) we arrive after some 
calculation to the following formula for the solution to (3.1)
\begin{align}
{\bm\xi}(t)=&\left(\frac{1}{3}+\frac{2}{3}\cos(t\sqrt{3}|{\bm a}|)\right){\bm\xi}_0
+\frac{2}{3|{\bm a}|}\sin(t\sqrt{3}|{\bm a}|){\bm a}\wedge{\bm\xi}_0\nonumber\\
&{}+(1-\cos(t\sqrt{3}|{\bm a}|))\left(-\frac{2}{3|{\bm a}|}{\bm a}*{\bm\xi}_0
+\frac{4}{3{\bm a}^2}({\bm a}\cdot{\bm\xi}_0){\bm a}\right).
\end{align}
Using (3.13) and (3.15) we find that the system (3.1) has in the discussed case the 
stationary solution representing the pure state of the form
\begin{equation}
\bar{\bm\xi}=\frac{{\bm a}}{|{\bm a}|}.
\end{equation}
Evidently, the stationary solution (3.20) is not asymptotically stable (this is 
not any limit of $t$ going to infinity of (3.19)).
Furthermore, the stationary solution representing the mixed state on the boundary 
of the state space is given by
\begin{equation}
\bar{\bm\xi}'=-\frac{{\bm a}}{2|{\bm a}|}.
\end{equation}
Finally, we have a family of stationary solutions corresponding to mixed states 
in the interior of the space of states such that
\begin{equation}
\bar{\bm\xi}''=-\frac{{\bm a}}{\sigma|{\bm a}|},
\end{equation}
where $\sigma>2$.  Of course, $\bar{\bm\xi}''$ approaches 0 referring to 
maximally mixed state in the limit $\sigma\to\infty$.

The projections of the periodic trajectory (3.19) onto the planes 
$(\xi_i,\xi_j)$, $i\neq j$, $i,j=1,2,\ldots,8$ are circles,  ellipses, segments 
and points so it is a plausible counterpart of the periodic evolution of the 
qubit.
\subsection{The case with ${\bm a}*{\bm a}=-|{\bm a}|{\bm a}$}
\label{sec:3.2}
It can be easily checked that the solution to (3.1) in the case of 
${\bm a}*{\bm a}=-|{\bm a}|{\bm a}$ can be obtained from the solution (3.19) to (3.1)
with ${\bm a}*{\bm a}=|{\bm a}|{\bm a}$ by the formal replacement ${\bm a}\to-{\bm a}$ 
and $t\to-t$.  Hence we get for ${\bm a}*{\bm a}=-|{\bm a}|{\bm a}$
\begin{align}
{\bm\xi}(t)=&\left(\frac{1}{3}+\frac{2}{3}\cos(t\sqrt{3}|{\bm a}|)\right){\bm\xi}_0
+\frac{2}{3|{\bm a}|}\sin(t\sqrt{3}|{\bm a}|){\bm a}\wedge{\bm\xi}_0\nonumber\\
&{}+(1-\cos(t\sqrt{3}|{\bm a}|))\left(\frac{2}{3|{\bm a}|}{\bm a}*{\bm\xi}_0
+\frac{4}{3{\bm a}^2}({\bm a}\cdot{\bm\xi}_0){\bm a}\right).
\end{align}
Analogously, applying the transformation ${\bm a}\to-{\bm a}$ to (3.20), (3.21) 
and (3.22) we arrive at the stationary solutions of the form
\begin{align}
\bar{\bm\xi} &=-\frac{{\bm a}}{|{\bm a}|}\qquad \text{(pure state)},\\
\bar{\bm\xi}'&=\frac{{\bm a}}{2|{\bm a}|}\qquad \text{(mixed state on the boundary of 
the state space)},\\
\bar{\bm\xi}''&=\frac{{\bm a}}{\sigma|{\bm a}|},\quad \sigma>2\quad \text{(mixed state in the 
interior of the space of states)}.
\end{align}
It is clear that qualitative behavior of solutions to (3.1) given by (3.23) is 
the same as in the case with ${\bm a}*{\bm a}=|{\bm a}|{\bm a}$.
\subsection{The condition ${\bm a}\cdot({\bm a}*{\bm a})=0$}
\label{sec:3.3}
Using the system of recurrence equations (3.3) for ${\bm a}\cdot({\bm a}*{\bm a})=0$, we easily find
\begin{equation}
e^{\tau{\bm a}\cdot{\bm\lambda}}=\frac{1}{3}+\frac{2}{3}\cosh(\tau|{\bm a}|)+\left[
\frac{\sinh(\tau|{\bm a}|)}{|{\bm a}|}{\bm a}+\frac{1}{\sqrt{3}{\bm a}^2}
(\cosh(\tau|{\bm a}|)-1){\bm a}*{\bm a}\right]\cdot{\bm\lambda}.
\end{equation}
On setting $\tau=-{\rm i}t$ and using (1.1), (2.7) and (2.8) we obtain after some calculation the 
solution to (3.1) such that
\begin{align}
{\bm\xi}(t)\nonumber\\
{}=&\frac{1}{3}(2\cos^2(t|{\bm a}|)+2\cos(t|{\bm a}|)-1){\bm\xi}_0+
\frac{2}{3\sqrt{3}|{\bm a}|}(2\cos(t|{\bm a}|)+1)\sin(t|{\bm a}|){\bm a}\wedge{\bm\xi}_0\nonumber\\
&{}+\frac{2}{3{\bm a}^2}(2\cos(t|{\bm a}|)+1)(\cos(t|{\bm a}|)-1)({\bm a}*{\bm a})*{\bm\xi}_0
+2\frac{\sin^2(t|{\bm a}|)}{{\bm a}^2}({\bm a}\cdot{\bm\xi}_0){\bm a}\nonumber\\
&{}+\frac{2}{3|{\bm a}|^4}(\cos(t|{\bm a}|)-1)^2[({\bm a}*{\bm a})\cdot{\bm\xi}_0]{\bm a}*{\bm a}
-\frac{2}{3\sqrt{3}|{\bm a}|^3}(\cos(t|{\bm a}|)-1)\sin(t|{\bm a}|)\nonumber\\
&\times\{{\bm a}*[({\bm a}*{\bm a})\wedge{\bm\xi}_0] -({\bm a}*{\bm a})*({\bm a}\wedge{\bm\xi}_0)\}.
\end{align}
\begin{figure*}
\centering
\includegraphics[width =.75\textwidth]{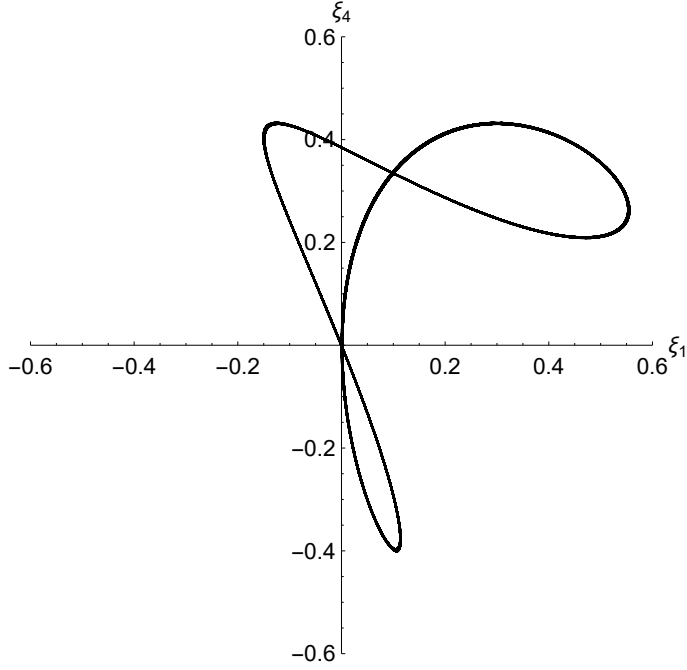}
\caption{The periodic solution of the system (3.1) corresponding to ${\bm a}\cdot({\bm a}*{\bm a})=0$, given by
(3.28), where ${\bm a}=(1,0,1,1,-1,1,1,0)$ and ${\bm\xi}_0=(0,0,\frac{\sqrt{3}}{2},0,0,0,0,\frac{1}{2})$.  The 
projection of the trajectory onto the $(\xi_1,\xi_4)$ plane.}
\end{figure*}

Now, (3.15) implies the following form of the stationary solutions to (3.1) referring 
to the pure states
\begin{equation}
\bar{\bm\xi}_1=-\frac{{\bm a}*{\bm a}}{{\bm a}^2},\quad \bar{\bm\xi}_{2,3}=\pm\frac{\sqrt{3}}{2|{\bm a}|}
{\bm a}+\frac{1}{2{\bm a}^2}{\bm a}*{\bm a}.
\end{equation}
Further, setting $\nu=0$ and ${\bm a}^2\mu^2+|{\bm a}|^4\nu^2=\frac{1}{3}$ in 
(3.16) with ${\bm a}\cdot({\bm a}*{\bm a})=0$, we obtain the following stationary 
solutions corresponding to the mixed states on the boundary of the space of states
\begin{equation}
\bar{\bm\xi}=\pm\frac{{\bm a}}{\sqrt{3}|{\bm a}|},\qquad (\nu=0)
\end{equation}
and
\begin{equation}
\bar{\bm\xi}_{1,2}=\frac{{\bm a}}{2\sqrt{3}|{\bm a}|}\pm\frac{{\bm a}*{\bm a}}{2{\bm a}^2},
\quad \bar{\bm\xi}_{3,4}=-\frac{{\bm a}}{2\sqrt{3}|{\bm a}|}\pm\frac{{\bm a}*{\bm 
a}}{2{\bm a}^2}\quad ({\bm a}^2\mu^2+|{\bm a}|^4\nu^2=\textstyle{\frac{1}{3})}
\end{equation}
Finally, assuming that ${\bm a}^2\mu^2+|{\bm a}|^4\nu^2=\textstyle{\frac{1}{3}}$ 
we find the stationary solutions (3.13) referring to mixed states in the interior of 
the space of states specified by
\begin{equation}
\mu=\pm\sqrt{\frac{1}{3}-|{\bm a}|^4\nu^2},\qquad 0<\nu<\frac{1}{2{\bm a}^2}.
\end{equation}
\subsection{Diagonal generator of evolution ${\bm a}\cdot{\bm\lambda}$}
We now discuss the case of the diagonal matrix ${\bm a}\cdot{\bm\lambda}$.  Since 
the only diagonal Gell-Mann matrices are $\lambda_3$ and $\lambda_8$, we then have
${\bm a}=(0,0,a_3,0,0,0,0,a_8)$.  We point out that since ${\bm a}*{\bm a}=
(0,0,2a_3a_8,0,0,0,0,a_3^2-a_8^2)$, the vector ${\bm a}$ satisfies the condition 
${\bm a}*{\bm a}=|{\bm a}|{\bm a}$ for $a_3$=0 and $a_8\leqslant0$, and for 
$a_3=\pm\sqrt{3}a_8$ and $a_8>0$, so we then have the case discussed in Sect.\ 3.1.
Analogously, for $a_3=0$ and $a_8\geqslant0$ as well as for 
$a_3=\pm\sqrt{3}a_8$ and $a_8<0$, we get the condition ${\bm a}*{\bm a}=-|{\bm a}|
{\bm a}$ and we then deal wit the case investigated in Sect.\ 3.2.  Furthermore, 
for $a_8=0$ and the condition $a_8\neq0$ and $a_3=\pm\frac{1}{\sqrt{3}}a_8$, we 
have ${\bm a}\cdot({\bm a}*{\bm a})=0$, so we then have the case analyzed in 
Sect.\ 3.3.

Now it can be easily checked that
\begin{equation}
e^{\tau(a_3\lambda_3+a_8\lambda_8)}=f(\tau)+\gamma(\tau)\lambda_3+\delta(\tau)\lambda_8,
\end{equation}
where
\begin{align}
f(\tau)&=\frac{1}{3}\left[e^{\tau(a_3+\frac{1}{\sqrt{3}})a_8)}+e^{\tau(-a_3+\frac{1}{\sqrt{3}})a_8)}
+e^{-2\tau\frac{a_8}{\sqrt{3}}}\right],\\
\gamma(\tau)&=\frac{1}{2}\left[e^{\tau(a_3+\frac{1}{\sqrt{3}})a_8)}-e^{\tau(-a_3+\frac{1}{\sqrt{3}})a_8)}\right],\\
\delta(\tau)&=\frac{\sqrt{3}}{6}\left[e^{\tau(a_3+\frac{1}{\sqrt{3}})a_8)}+e^{\tau(-a_3+\frac{1}{\sqrt{3}})a_8)}
-2e^{-2\tau\frac{a_8}{\sqrt{3}}}\right].
\end{align}
On setting $\tau=-{\rm i}t$  and maing use of (1.1), (2.7) and (2.8) we arrive after some 
calculation at the following solution of (3.1)
\begin{equation}
\begin{split}
\xi_1(t) &= \cos(2ta_3)\xi_{01}-\sin(2ta_3)\xi_{02},\\
\xi_2(t) &= \sin(2ta_3)\xi_{01}+\cos(2ta_3)\xi_{02},\\
\xi_3(t) &= \xi_{03},\\
\xi_4(t) &= \cos[t(a_3+\sqrt{3}a_8)]\xi_{04}-\sin[t(a_3+\sqrt{3}a_8)]\xi_{05},\\
\xi_5(t) &= \sin[t(a_3+\sqrt{3}a_8)]\xi_{04}+\cos[t(a_3+\sqrt{3}a_8)]\xi_{05},\\ 
\xi_6(t) &= \cos[t(a_3-\sqrt{3}a_8)]\xi_{06}+\sin[t(a_3-\sqrt{3}a_8)]\xi_{07},\\
\xi_7(t) &= -\sin[t(a_3-\sqrt{3}a_8)]\xi_{06}+\cos[t(a_3-\sqrt{3}a_8)]\xi_{07},\\
\xi_8(t) &= \xi_{08}. 
\end{split}
\end{equation}
The solution (3.37) is in general three-frequency quasi-periodic solution.  This 
observation is consisitent with the general formula (3.12) for the exponential of 
${\bm a}\cdot{\bm\lambda}$.  Examples of the trajectories given by (3.37) that are 
the Lissajous curves are illustrated in Fig.\ 3.
\begin{figure*}
\centering
\begin{tabular}{c@{}c}
\includegraphics[width =.49\textwidth]{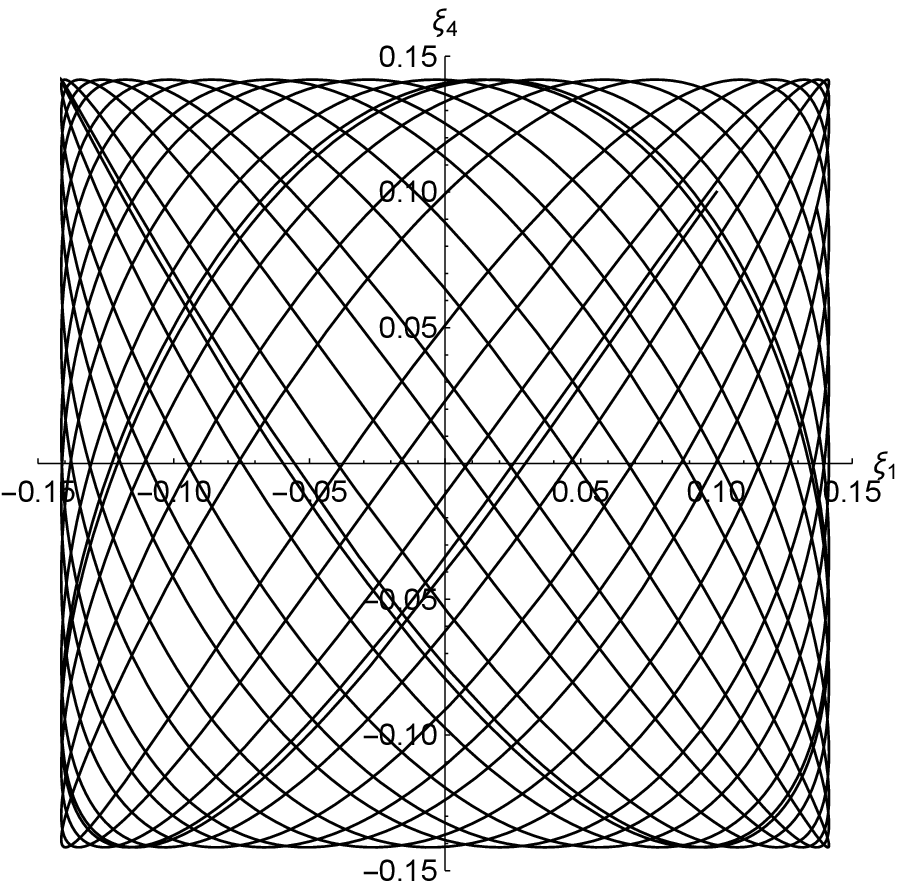}&
\includegraphics[width =.49\textwidth]{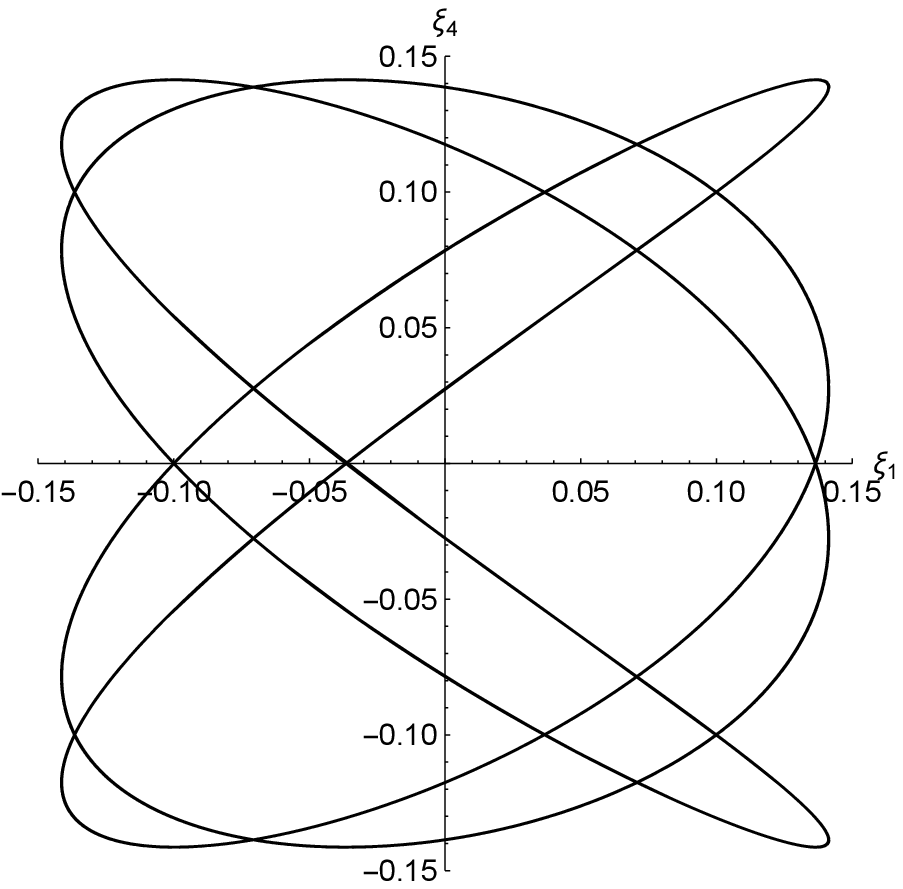}
\end{tabular}
\caption{Trajectories of the system (3.1) given by (3.37) corresponding to the diagonal 
operator of evolution.  Left: the quasiperiodic trajectory specified by
${\bm a}=(0,0,\frac{1}{2},0,0,0,0,\frac{1}{2})$ and 
${\bm\xi}_0=(0.1,0.1,0.1,0.1,0.1,0.1,0.1,0.1)$  for $t=100$.  The projection of the trajectory
onto the $(\xi_1,\xi_4)$ plane. Right: the periodic trajectory referring to 
${\bm a}=(0,0,\sqrt{3},0,0,0,0,\frac{1}{2})$ and ${\bm\xi}_0=(0.1,0.1,0.1,0.1,0.1,0.1,0.1,0.1)$.}
\end{figure*}

We now discuss the stationary solutions.  
From (3.13) and (A.10) it follows that the stationary solutions in the discussed case are of 
the form $\bar{\bm\xi}=(0,0,\bar\xi_3,0,0,0,0,\bar\xi_8)$.  Hence we find that 
the stationary solutions referring to the pure states are 
$\bar{\bm\xi_1}~=~(0,0,0,0,0,0,0,-1)$, and $\bar{\bm\xi}_{2,3}~=~(0,0,\pm\frac{\sqrt{3}}{2},0,0,0,0,
\frac{1}{2})$.  The stationary solutions corresponding to the mixed states on the 
boundary of the state space can be easily obtained from (2.5).  Namely, we have
\begin{align}
\bar\xi_8=&\frac{1}{2}\sqrt[3]{\frac{1}{2}(1-3\kappa)+\sqrt{-\kappa^3+\frac{1}{4}(1-3\kappa)^2}}\nonumber\\
&{}+\frac{1}{2}\sqrt[3]{\frac{1}{2}(1-3\kappa)-\sqrt{-\kappa^3+\frac{1}{4}(1-3\kappa)^2}}\\
\bar\xi_{3;1,2}=&\pm\sqrt{\kappa-{\bar\xi_8}^2},\qquad\qquad\qquad \kappa\in\left(0,\frac{1}{4}\right],\nonumber
\end{align}
where $\kappa$ is the parameter defined by $\bar\xi_3^2+\bar\xi_8^2=\kappa$, so 
in the general case of the mixed states on the boundary it satisfies 
$\kappa\in(0,1)$.  In the particular case of $\kappa=\frac{1}{4}$ the relations 
(3.38) reduce to $\bar\xi_8=\frac{1}{2}$ and $\bar\xi_3=0$.  Further, the 
trigonometric solution to the cubic equation satisfied by $\bar\xi_8$ results in
\begin{equation}
\begin{split}
\bar\xi_{8,1}=&\sqrt{\kappa}\cos\frac{\alpha}{3},\qquad 
\bar\xi_{8;2,3}=-\sqrt{\kappa}\cos\left(\frac{\alpha}{3}\pm\frac{\pi}{3}\right),\\
\bar\xi_{3;1,2}=&\pm\sqrt{\kappa-{\bar\xi_8}^2},\qquad\qquad\qquad\kappa\in\left(\frac{1}{4},1\right),
\end{split}
\end{equation}
where $\cos\alpha=-4\frac{3\kappa-1}{\sqrt{\kappa^3}}$.  For $\kappa=\frac{1}{3}$ 
the formulas (3.39) simplify to $\bar\xi_{8,1}=\frac{1}{2}$, $\bar\xi_{8,2}=0$, $\bar\xi_{8,3}=-\frac{1}{2}$,
corresponding to $\bar\xi_{3;1,3}=\pm\frac{1}{2\sqrt{3}}$, $\bar\xi_{3,2}=\pm\frac{1}{\sqrt{3}}$, respectively.
The stationary solutions referring to the mixed states in the interior of the state space are specified by the
system of the inequalities such that (see (2.2)) 
\begin{equation}
\begin{cases}
\bar\xi_3^2+\bar\xi_8^2<1,\\
2\bar\xi_8^3-6\bar\xi_3^2\bar\xi_8+3(\bar\xi_3^2+\bar\xi_8^2)<1.
\end{cases}
\end{equation}
The region in the plane defined by (3.40) is depicted in Fig.\ 4. The vertices of the 
triangle correspond to the pure states and its edges refer to the mixed states on 
the boundary of the space of states.  The parametrization of edges is given by (3.38) 
and (3.39).  We finally point out that the case of the diagonal Gell-Mann matrices was
analyzed in the context of the properties of the states of qutrits in \cite{16,17}.
\begin{figure*}
\centering
\includegraphics[width =.75\textwidth]{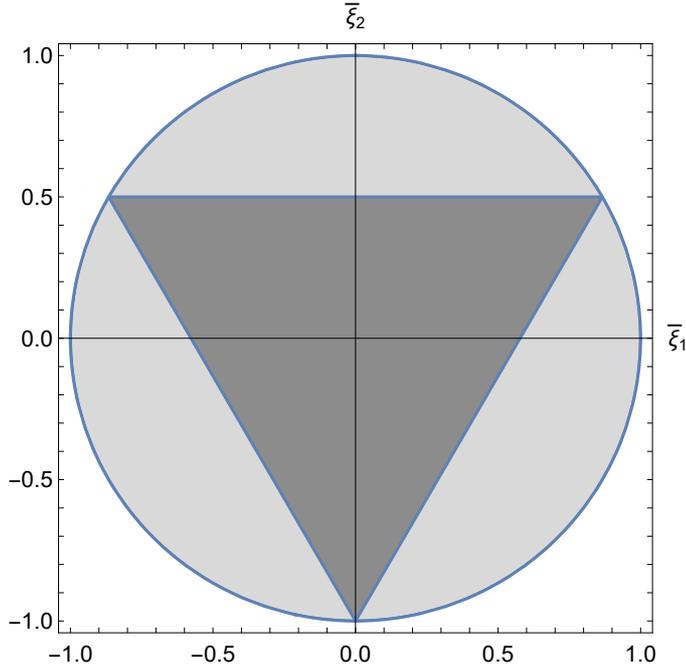}
\caption{The region in the plane (darker triangle) containing stationary solutions of (3.1)
corresponding to mixed states in the interior of the state space, given by (3.40) (the case of
the diagonal generator of evolution).  The vertices of the triangle $(-\frac{\sqrt{3}}{2},\frac{1}{2})$, 
$(0,-1)$, and $(\frac{\sqrt{3}}{2},\frac{1}{2})$ refer to pure states and its edges to mixed states
on the boundary of the space of states.}
\end{figure*}

\section{Nonlinear evolution of a qutrit}
\label{sec:4}
We first confine ourselves to the case of ${\bm a}=0$, i.e.\ $H=0$.  The 
system (2.9) takes then the form
\begin{equation}
\dot{\bm\xi}=\frac{2}{\sqrt{3}}{\bm b}+\frac{2}{\sqrt{3}}{\bm b}*{\bm \xi}-\frac{4}{\sqrt{3}}
({\bm b}\cdot{\bm\xi}){\bm\xi},\qquad {\bm\xi}(0)={\bm\xi}_0.
\end{equation}
\subsection{The condition ${\bm b}*{\bm b}=|{\bm b}|{\bm b}$}
\label{sec:4.1}
Proceeding analogously as with (3.1) we first consider the case with ${\bm b}*{\bm b}=|{\bm b}|
{\bm b}$.  Taking into account (3.18) with $\tau=t$, (1.1), (2.7) and (2.8), we get the solution 
to (4.1) of the form
\begin{equation}
{\bm \xi}(t)=\frac{{\bm\eta}(t)}{\varphi(t)},
\end{equation}
where
\begin{align}
{\bm\eta}(t)=& \left(-\frac{1}{9}e^{4t\frac{\sqrt{3}}{3}|{\bm b}|}+\frac{8}{9}e^{t\frac{\sqrt{3}}{3}|{\bm b}|}
+\frac{2}{9}e^{-2t\frac{\sqrt{3}}{3}|{\bm b}|}\right){\bm\xi}_0\nonumber\\
&{}+\frac{1}{9|{\bm b}|}\left(e^{4t\frac{\sqrt{3}}{3}|{\bm b}|}+4e^{t\frac{\sqrt{3}}{3}|{\bm b}|}
-5e^{-2t\frac{\sqrt{3}}{3}|{\bm b}|}\right){\bm b}*{\bm\xi}_0\nonumber\\
&{}+\frac{4}{9{\bm b}^2}\left(e^{4t\frac{\sqrt{3}}{3}|{\bm b}|}-2e^{t\frac{\sqrt{3}}{3}|{\bm b}|}
+e^{-2t\frac{\sqrt{3}}{3}|{\bm b}|}\right)({\bm b}\cdot{\bm\xi}_0){\bm b}\nonumber\\
&{}+\frac{2}{9{\bm b}^2}\left(e^{4t\frac{\sqrt{3}}{3}|{\bm b}|}-2e^{t\frac{\sqrt{3}}{3}|{\bm b}|}
+e^{-2t\frac{\sqrt{3}}{3}|{\bm b}|}\right)({\bm b}*{\bm\xi}_0)*{\bm b}\nonumber\\
&{}+\frac{1}{3|{\bm b}|}\left(e^{4t\frac{\sqrt{3}}{3}|{\bm b}|}-e^{-2t\frac{\sqrt{3}}{3}|{\bm b}|}\right){\bm b},
\end{align}
\begin{equation}
\varphi(t)=\frac{1}{3}e^{4t\frac{\sqrt{3}}{3}|{\bm b}|}+\frac{2}{3}e^{-2t\frac{\sqrt{3}}{3}|{\bm b}|}
+\frac{2}{3|{\bm b}|}\left(e^{4t\frac{\sqrt{3}}{3}|{\bm 
b}|}-e^{-2t\frac{\sqrt{3}}{3}|{\bm b}|}\right){\bm b}\cdot{\bm\xi}_0.
\end{equation}
Taking into account (4.1) one can easily check that we have in the discussed case the following 
stationary solution
\begin{equation}
\bar{\bm\xi}=\frac{{\bm b}}{|{\bm b}|}.
\end{equation}
It follows from numerical calculations that the stationary solution (4.5) is 
asymptotically stable.  An easy example of solution to (4.1) going to 
$\bar{\bm\xi}$ as $t\to\infty$ can be obtained from (4.4) by setting 
${\bm\xi}_0=0$.  The problem of finding the stationary solutions to (4.1) that are not
asymptotically stable seems to be a difficult task.  Such solution proportional to ${\bm b}$ is of the form
\begin{equation}
\bar{\bm\xi}'=-\frac{{\bm b}}{2|{\bm b}|}.
\end{equation} 
It can be easily verified that (4.5) and (4.6) represent the pure state and the 
mixed state on the boundary of the state space, respectively.
\subsection{The case of ${\bm b}*{\bm b}=-|{\bm b}|{\bm b}$}
\label{sec:4.2}
Proceeding analogously as in Sect.\ 3.2 and replacing in (4.3) ${\bm b}$ with 
$-{\bm b}$ and $t$ with $-t$ we find that the solution to (4.1) corresponding to
${\bm b}*{\bm b}=-|{\bm b}|{\bm b}$ is given by (4.2), where
\begin{align}
{\bm\eta}(t)=& \left(\frac{2}{9}e^{2t\frac{\sqrt{3}}{3}|{\bm b}|}+\frac{8}{9}e^{-t\frac{\sqrt{3}}{3}|{\bm b}|}
-\frac{1}{9}e^{-4t\frac{\sqrt{3}}{3}|{\bm b}|}\right){\bm\xi}_0\nonumber\\
&{}+\frac{1}{9|{\bm b}|}\left(5e^{2t\frac{\sqrt{3}}{3}|{\bm b}|}-4e^{-t\frac{\sqrt{3}}{3}|{\bm b}|}
-e^{-4t\frac{\sqrt{3}}{3}|{\bm b}|}\right){\bm b}*{\bm\xi}_0\nonumber\\
&{}+\frac{4}{9{\bm b}^2}\left(e^{2t\frac{\sqrt{3}}{3}|{\bm b}|}-2e^{-t\frac{\sqrt{3}}{3}|{\bm b}|}
+e^{-4t\frac{\sqrt{3}}{3}|{\bm b}|}\right)({\bm b}\cdot{\bm\xi}_0){\bm b}\nonumber\\
&{}+\frac{2}{9{\bm b}^2}\left(e^{2t\frac{\sqrt{3}}{3}|{\bm b}|}-2e^{-t\frac{\sqrt{3}}{3}|{\bm b}|}
+e^{-4t\frac{\sqrt{3}}{3}|{\bm b}|}\right)({\bm b}*{\bm\xi}_0)*{\bm b}\nonumber\\
&{}+\frac{1}{3|{\bm b}|}\left(e^{2t\frac{\sqrt{3}}{3}|{\bm b}|}-e^{-4t\frac{\sqrt{3}}{3}|{\bm b}|}\right){\bm b},
\end{align}
\begin{equation}
\varphi(t)=\frac{2}{3}e^{2t\frac{\sqrt{3}}{3}|{\bm b}|}+\frac{1}{3}e^{-4t\frac{\sqrt{3}}{3}|{\bm b}|}
+\frac{2}{3|{\bm b}|}\left(e^{2t\frac{\sqrt{3}}{3}|{\bm 
b}|}-e^{-4t\frac{\sqrt{3}}{3}|{\bm b}|}\right){\bm b}\cdot{\bm\xi}_0.
\end{equation}
The system (4.1) has in the discussed case the stationary solution such that
\begin{equation}
\bar{\bm\xi}=\frac{{\bm b}}{2|{\bm b}|}.
\end{equation}
The numerical simulations indicate that the stationary solution (4.9) is asymptotically stable.
The solution (4.9) represents the mixed state on the boundary of the space of 
states. We point out that there are no asymptotically stable stationary solutions 
representing mixed states in the case of the qubit \cite{1}. The asymptotically stable stationary 
solution (4.9) is not globally stable.  Consider for example the case of ${\bm b}=(0,0,\frac{\sqrt{3}}{2},
0,0,0,0,-\frac{1}{2})$.  It appears that the mixed states on the boundary of the state space 
${\bm\xi}_0=(0,0,\frac{1}{2},0,0,0,0,\frac{1}{2})$ and ${\bm\xi}_0=(0,0,\frac{1}{2}(1-\sqrt{3}),0,0,0,0,
\frac{1}{2}(1-\sqrt{3}))$ approach the stationary solutions representing the pure states
$\bar{\bm\xi}=(0,0,\frac{\sqrt{3}}{2},0,0,0,0,-\frac{1}{2})$ and 
$\bar{\bm\xi}=(0,0,0,0,0,0,0,-1)$, respectively.  Furthermore, we have the stationary solution 
to (4.1) that is not asymptotically stable of the form
\begin{equation}
\bar{\bm\xi}'=-\frac{{\bm b}}{|{\bm b}|}.
\end{equation}
The solution (4.10) represents the pure state.  It is clear that it is not the unique one.
\subsection{The condition ${\bm b}\cdot({\bm b}*{\bm b})=0$}
\label{sec:4.3}
We now discuss the case that is the nonlinear counterpart of the condition 
investigated in Sect.\ 3.3.  Applying the algorithm used for calculation of (3.28) 
we arrive at the following solution to (4.1) given by (4.2) and the relations
\begin{align}
&{\bm\eta}(t)\nonumber\\
&{}=\frac{1}{3}(2\cosh(t|{\bm b}|)+1){\bm\xi}_0+\frac{2}{\sqrt{3}}\frac{\sinh(t|{\bm b}|)}{|{\bm b}|}
{\bm b}*{\bm\xi}_0+\frac{2}{3{\bm b}^2}(1-\cosh(t|{\bm b}|)({\bm b}*{\bm b})*{\bm\xi}_0\nonumber\\
&{}+2\frac{\sinh(t|{\bm b}|)}{|{\bm b}|}\left[\frac{\sinh(t|{\bm b}|)}{|{\bm b}|}{\bm b}\cdot{\bm\xi}_0
+\frac{1}{\sqrt{3}{\bm b}^2}(\cosh(t|{\bm b}|)-1)({\bm b}*{\bm b})\cdot{\bm\xi}_0\right]{\bm b}\nonumber\\
&{}+\frac{2}{\sqrt{3}{\bm b}^2}(\cosh(t|{\bm b}|)-1)\left[\frac{\sinh(t|{\bm b}|)}{|{\bm b}|}{\bm b}\cdot{\bm\xi}_0
+\frac{1}{\sqrt{3}{\bm b}^2}(\cosh(t|{\bm b}|)-1)({\bm b}*{\bm b})\cdot{\bm\xi}_0\right]\nonumber\\
&\times{\bm b}*{\bm b}\nonumber+\frac{1}{\sqrt{3}}\frac{\sinh(2t|{\bm b}|)}{|{\bm b}|}{\bm b}+
\frac{1}{3{\bm b}^2}(\cosh(2t|{\bm b}|)-1){\bm b}*{\bm b},\\
\end{align}
\begin{equation}
\varphi(t)=\frac{2}{3}\cosh(2t|{\bm b}|)+\frac{1}{3}+\frac{2\sqrt{3}}{3}\frac{\sinh(2t|{\bm b}|)}{|{\bm b}|}{\bm b}
\cdot{\bm\xi}_0+\frac{2}{3{\bm b}^2}(\cosh(2t|{\bm b}|)-1)({\bm b}*{\bm b})\cdot{\bm\xi}_0.
\end{equation}
On making the ansatz of the form (3.13) with ${\bm a}$ replaced by ${\bm b}$ one 
can easily obtain the following stationary solutions to (4.1)
\begin{align}
\bar{\bm\xi}&=\frac{\sqrt{3}}{2|{\bm b}|}{\bm b}+\frac{{\bm b}*{\bm b}}{2{\bm b}^2},\\
\bar{\bm\xi}'&=-\frac{\sqrt{3}}{2|{\bm b}|}{\bm b}+\frac{{\bm b}*{\bm b}}{2{\bm b}^2},\\
\bar{\bm\xi}''&=-\frac{{\bm b}*{\bm b}}{{\bm b}^2}.
\end{align}
All the stationary solutions represent the pure states.  The numerical 
calculations show that the solution $\bar{\bm\xi}$ is asymptotically stable and 
the solutions $\bar{\bm\xi}'$ and $\bar{\bm\xi}''$ are unstable.
\subsection{Diagonal matrix ${\bm b}\cdot{\bm\lambda}$}
\label{sec:4.4}
We now study the case of ${\bm b}\cdot{\bm\lambda}=b_3\lambda_3+b_8\lambda_8$, 
that is the counterpart of the diagonal generator of the unitary evolution 
analyzed in Sect.\ 3.4.  Using (3.33) with $\tau$ replaced by $t$ we find after 
some algebra the following solution to (4.1)
\begin{equation}
\begin{split}
\eta_1(t) =& e^{2t\frac{b_8}{\sqrt{3}}}\xi_{01},\\
\eta_2(t) =& e^{2t\frac{b_8}{\sqrt{3}}}\xi_{02},\\
\eta_3(t) =& \frac{1}{2}\left[e^{2t(b_3+\frac{1}{\sqrt{3}}b_8)}+e^{2t(-b_3+\frac{1}
{\sqrt{3}}b_8)}\right]\xi_{03}+f_{38}(t)\xi_{08}\\
&{}+\frac{1}{2\sqrt{3}}\left[e^{2t(b_3+\frac{1}{\sqrt{3}}b_8)}-e^{2t(-b_3+\frac{1}{\sqrt{3}}b_8)}\right],\\
\eta_4(t) =& f_{45}(t)\xi_{04},\\
\eta_5(t) =& f_{45}(t)\xi_{05},\\
\eta_6(t) =& f_{67}(t)\xi_{06},\\
\eta_7(t) =& f_{67}(t)\xi_{07},\\
\eta_8(t) =& f_{38}(t)\xi_{03}+\left\{\frac{1}{6}\left[e^{2t(b_3+\frac{1}{\sqrt{3}}b_8)}+e^{2t(-b_3+\frac{1}
{\sqrt{3}}b_8)}\right]+\frac{2}{3}e^{-4t\frac{b_8}{\sqrt{3}}}\right\}\xi_{08}\\
&{}+\frac{1}{6}\left[e^{2t(b_3+\frac{1}{\sqrt{3}}b_8)}+e^{2t(-b_3+\frac{1}{\sqrt{3}}b_8)}
-2e^{-4t\frac{b_8}{\sqrt{3}}}\right],
\end{split}
\end{equation}
where
\begin{align}
f_{38}(t) =& \frac{1}{9}\left[(2\sqrt{3}-1)e^{2t(b_3+\frac{1}{\sqrt{3}}b_8)}-(2\sqrt{3}+1)e^{2t(-b_3+\frac{1}
{\sqrt{3}}b_8)}\right.\nonumber\\
&{}\left.+(-\sqrt{3}+1)e^{t(b_3-\frac{1}{\sqrt{3}}b_8)}+(\sqrt{3}+1)e^{t(-b_3-\frac{1}{\sqrt{3}}b_8)}
+e^{2t\frac{1}{\sqrt{3}}b_8}-e^{-4t\frac{1}{\sqrt{3}}b_8}\right],
\end{align}
\begin{align}
f_{45}(t) =& \left(\frac{1}{12}-\frac{\sqrt{3}}{18}\right)e^{2t(b_3+\frac{1}{\sqrt{3}}b_8)}-
\left(\frac{1}{12}+\frac{\sqrt{3}}{18}\right)e^{2t(-b_3+\frac{1}{\sqrt{3}}b_8)}\nonumber\\
&{}+\left(\frac{5}{6}+\frac{\sqrt{3}}{18}\right)e^{t(b_3-\frac{1}{\sqrt{3}}b_8)}+
\left(\frac{1}{6}+\frac{\sqrt{3}}{18}\right)e^{t(-b_3-\frac{1}{\sqrt{3}}b_8)}\nonumber\\
&{}+\frac{\sqrt{3}}{18}\left(e^{2t\frac{1}{\sqrt{3}}b_8}-e^{-4t\frac{1}{\sqrt{3}}b_8}\right),
\end{align}
\begin{align}
f_{67}(t) =& 
\left(-\frac{1}{12}+\frac{\sqrt{3}}{18}\right)e^{2t(b_3+\frac{1}{\sqrt{3}}b_8)}+
\left(\frac{1}{12}+\frac{\sqrt{3}}{18}\right)e^{2t(-b_3+\frac{1}{\sqrt{3}}b_8)}\nonumber\\
&{}+\left(\frac{1}{6}-\frac{\sqrt{3}}{18}\right)e^{t(b_3-\frac{1}{\sqrt{3}}b_8)}+
\left(\frac{5}{6}-\frac{\sqrt{3}}{18}\right)e^{t(-b_3-\frac{1}{\sqrt{3}}b_8)}\nonumber\\
&{}-\frac{\sqrt{3}}{18}\left(e^{2t\frac{1}{\sqrt{3}}b_8}-e^{-4t\frac{1}{\sqrt{3}}b_8}\right).
\end{align}
\begin{align}
\varphi(t)=&\frac{1}{3}\left(e^{2t(b_3+\frac{1}{\sqrt{3}}b_8)}+e^{2t(-b_3+\frac{1}{\sqrt{3}}b_8)}
+e^{-4t\frac{1}{\sqrt{3}}b_8}\right)\nonumber\\
&{}+\frac{\sqrt{3}}{3}\left(e^{2t(b_3+\frac{1}{\sqrt{3}}b_8)}-e^{2t(-b_3+\frac{1}{\sqrt{3}}b_8)}\right)
\xi_{03}\nonumber\\
&{}+\frac{1}{3}\left(e^{2t(b_3+\frac{1}{\sqrt{3}}b_8)}+e^{2t(-b_3+\frac{1}{\sqrt{3}}b_8)}
-2e^{-4t\frac{1}{\sqrt{3}}b_8}\right)\xi_{08}.
\end{align}
As mentioned earlier in Sect.\ 3.4 $b_3=\pm\sqrt{3}b_8$ and $b_8>0$ correspond 
to ${\bm b}*{\bm b}=|{\bm b}|{\bm b}$; $b_3=0$ and $b_8\geqslant0$ as well as 
$b_3=\pm\sqrt{3}b_8$ and $b_8<0$ refer to ${\bm b}*{\bm b}=-|{\bm b}|{\bm b}$.  
Furthermore, $b_8=0$ and the conditions $b_3=\pm\frac{1}{\sqrt{3}}a_8$ and 
$b_8\neq0$ lead to ${\bm b}\cdot({\bm b}*{\bm b})=0$.   Of course, in such cases 
we can apply the results of the previous sections concerning stationary solutions.  
The new stationary solutions can be obtained for example by setting $\bar{\bm\xi}=
(0,0,\bar{\xi}_3,0,0,0,0,\bar{\xi}_8)$.  Namely, we find 
$\bar{\bm\xi}_{1,2}=(0,0,\pm\frac{\sqrt{3}}{2},0,0,0,0,\frac{1}{2})$ and 
$\bar{\bm\xi}_3=(0,0,0,0,0,0,0,-1)$.  This solutions represent the pure states.  
It follows from numerical calculations that the solution $\bar{\bm\xi}_1$ is 
asymptotically stable and $\bar{\bm\xi}_2$ and $\bar{\bm\xi}_3$ are unstable.
\subsection{Rational solution}
\label{sec:4.5}
We finally discuss the case when the series representing the exponential from (1.1) 
such that $e^{-{\rm i}t(H+{\rm i}G)}=e^{-{\rm i}t({\bm a}+{\rm i}{\bm b})\cdot{\bm\lambda}}$ 
truncates and becomes a polynomial, so the solution to (2.9) is rational.  This is the 
only simple 
example involving nonvanishing both ${\bm a}$ and ${\bm b}$.  We recall that in 
the case of the qubit the rational solution was intermediate between the periodic 
and hyperbolic motion.  Taking into account (A.31) we find that whenever ${\bm a}^2=
{\bm b}^2$, ${\bm a}\cdot{\bm b}=0$, and ${\bm a}*{\bm a}={\bm b}*{\bm b}$, 
${\bm a}*{\bm b}=0$, then
\begin{equation}
e^{-{\rm i}t({\bm a}+{\rm i}{\bm b})\cdot{\bm\lambda}}=1-{\rm i}t({\bm a}+{\rm i}{\bm b})
\cdot{\bm\lambda}.
\end{equation}
We remark that the conditions satisfied by ${\bm a}$ and ${\bm b}$ in the case of 
the rational solution are the most natural generalizations of those taking place 
for the qubit such that ${\bm a}^2={\bm b}^2$ and ${\bm a}\cdot{\bm b}=0$.  It 
should also be noted that relations satisfied by ${\bm a}$ and ${\bm b}$ imply 
${\bm a}\cdot({\bm a}*{\bm a})=0$ and ${\bm b}\cdot({\bm b}*{\bm b})=0$ following 
directly from (A.27).  Now, making use of (4.21) we get after some calculation the 
following solution to (2.9) expressed by (4.2) with
\begin{align}
&{\bm\eta}(t) = {\bm\xi}_0+\frac{2}{\sqrt{3}}({\bm b}+{\bm b}*{\bm\xi}_0 +{\bm a}
\wedge{\bm\xi}_0)t+\left\{\frac{2}{3}{\bm a}*{\bm a}+\frac{2}{3}{\bm a}\wedge{\bm b}
-\frac{2}{3}{\bm a}^2{\bm\xi}_0\right.\nonumber\\
&\left.{}-\frac{4}{3}({\bm a}*{\bm a})*{\bm\xi}_0+2({\bm a}\cdot{\bm\xi}_0){\bm a}+
2({\bm b}\cdot{\bm\xi}_0){\bm b}-2[{\bm a}*({\bm b}\wedge{\bm\xi}_0)-{\bm a}\wedge
({\bm b}*{\bm\xi}_0)]\right\}t^2,
\end{align}
\begin{equation}
\varphi(t)=1+\frac{4\sqrt{3}}{3}{\bm b}\cdot{\bm\xi}_0t+\frac{4}{3}[{\bm a}^2
+({\bm a}*{\bm a})\cdot{\bm\xi}_0-({\bm a}\wedge{\bm b})\cdot{\bm\xi}_0]t^2.
\end{equation}
On putting ${\bm\xi}_0=0$ and taking the limit $\lim_{t\to\infty}\frac{{\bm\eta}(t)}
{\varphi(t)}$, we arrive at the asymptotically stable stationary solution of the 
form
\begin{equation}
\bar{\bm\xi}=\frac{1}{2}\frac{{\bm a}*{\bm a}}{{\bm a}^2}+\frac{1}{2}\frac{{\bm a}
\wedge{\bm b}}{{\bm a}^2}.
\end{equation}
This stationary solution satisfies ${\bm\xi}^2=\frac{1}{2}$ and $3{\bm\xi}^2-2\bar{\bm\xi}
\cdot(\bar{\bm\xi}*\bar{\bm\xi})=1$, so it represents a mixed state on the 
boundary of the space of states.
\subsection{Limit cycle, spiral trajectories and remaining solutions for ${\bm a}\neq0$ and 
${\bm b}\neq0$}
\label{sec:4.6}
An interesting nontrivial property of the nonlinear system (2.9) absent in the 
case of the qubit, is the existence of the limit cycles.  More precisely, it 
follows from the numerical calculations that whenever ${\bm a}^2>{\bm b}^2$, ${\bm a}\cdot{\bm b}=0$,
${\bm a}\cdot({\bm a}*{\bm a})=0$, ${\bm a}\cdot({\bm b}*{\bm b})=0$, and ${\bm b}\cdot({\bm a}*{\bm a})>0$,
then the trajectories go to the periodic solution represented in the projections 
onto the planes $(\xi_i,\xi_j)$ or $(\xi_i,\xi_j,\xi_k)$, $i,j,k=1,2,\ldots,8$ by 
ellipses.  We point out that the first two conditions ${\bm a}^2>{\bm b}^2$ and 
${\bm a}\cdot{\bm b}=0$ refer to periodic solutions in the case of the nonlinear 
evolution of the qubit \cite{1}.  The limit cycle of the system (2.9) is 
illustrated in the Fig.\ 5.  It should be noted that the parameters of the limit 
cycles depend on initial data. 
\begin{figure*}
\centering
\includegraphics[width =.75\textwidth]{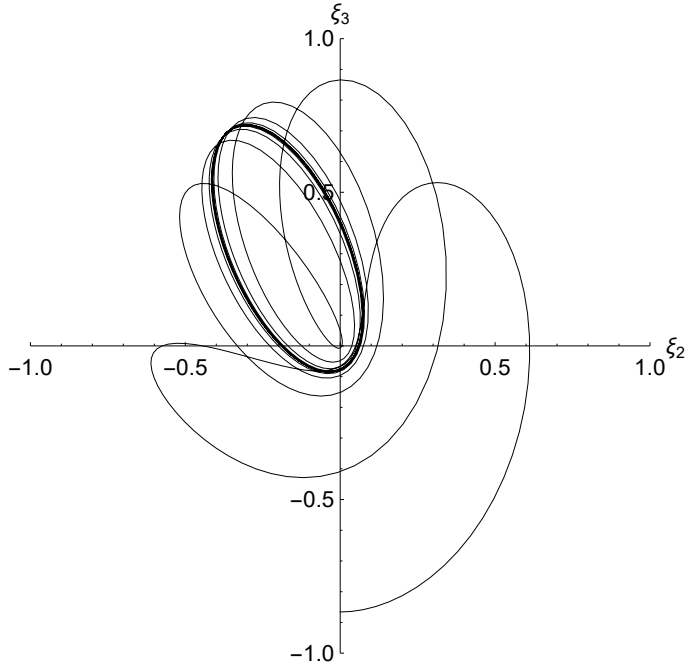}
\caption{The limit cycle of the system (2.9), where ${\bm a}=(1,1,0,2,-2,1,0,0)$, 
${\bm b}=(0,0,\frac{\sqrt{3}}{2},0,0,0,0,\frac{1}{2})$ and
${\bm\xi}_0=(0,0,-\frac{\sqrt{3}}{2},0,0,0,0,\frac{1}{2})$.}
\end{figure*}
As with the nonlinear evolution of the qubit there 
exist for ${\bm a}\cdot{\bm b}\neq0$ the spiral trajectories going to equilibrium 
points (stationary solutions) corresponding to the pure states that are 
combinations of periodic and hyperbolic motion.  An example is of the form 
${\bm a}=(1,1,0,2,-2,1,0,1)$, ${\bm b}=(0,0,\frac{\sqrt{3}}{2},0,0,0,0,-\frac{1}{2})$
and ${\bm\xi}_0=(0.1,0.1,0.1,0.1,0.1,0.1,0.1,0.1)$.  
The explicit solution for ${\bm a}\cdot{\bm b}\neq0$
can be easily obtained by setting ${\bm a}=\mu{\bm b}$ and using solutions obtained earlier.
For instance, one can demand that ${\bm b}\cdot({\bm b}*{\bm b})=0$ and use solutions (3.28) 
and (4.11).  In opposition to the case of the qubit we have also the spiral 
solutions corresponding to ${\bm a}\cdot{\bm b}=0$.  The conditions for 
parameters can be then obtained from that holding in the case of the limit cycle by 
omitting some of requirements.  An example is ${\bm a}=(0,1,0,2,-2,1,0,0)$ and 
${\bm b}=(0,0,\frac{\sqrt{3}}{2},0,0,0,0,-\frac{1}{2})$ satisfying ${\bm a}\cdot({\bm a}*{\bm a})\neq0$
as well as ${\bm a}=(1,1,0,2,-2,1,0,0)$ and ${\bm b}=(0,0,\frac{1}{2}(1-\sqrt{3}),0,0,0,0,\frac{1}{2}(1-\sqrt{3})$
fulfilling ${\bm a}\cdot({\bm b}*{\bm b})\neq0$ and ${\bm b}\cdot({\bm a}*{\bm a})<0$. 
In both examples ${\bm\xi}_0=(0.1,0.1,0.1,0.1,0.1,0.1,0.1,0.1)$.  An interesting new type of solutions
that occur in the discussed case of the qutrit are dumped quasi-periodic 
oscillations such that the volume of the phase space occupied by the trajectory 
shrinks to zero and solution goes to the equilibrium point representing a pure 
state.  Such oscillations that can be regarded as the quasi-periodic counterpart 
of the spiral motion are depicted in Fig.\ 6.  We finally point out that the 
Gell-Mann matrices $\lambda_1$, $\lambda_2$ and $\lambda_3$ are generators of the 
$SU(2)$ group, so we can recover all solutions obtained in the case of the qubit 
\cite{1} by putting ${\bm a}=(a_1,a_2,a_3,0,0,0,0,0)$, ${\bm b}=(b_1,b_2,b_3,0,0,0,0,0)$
and ${\bm\xi}_0=(\xi_{01},\xi_{02},\xi_{03},0,0,0,0,0)$.
\begin{figure}
\centering
\includegraphics[width =.75\textwidth]{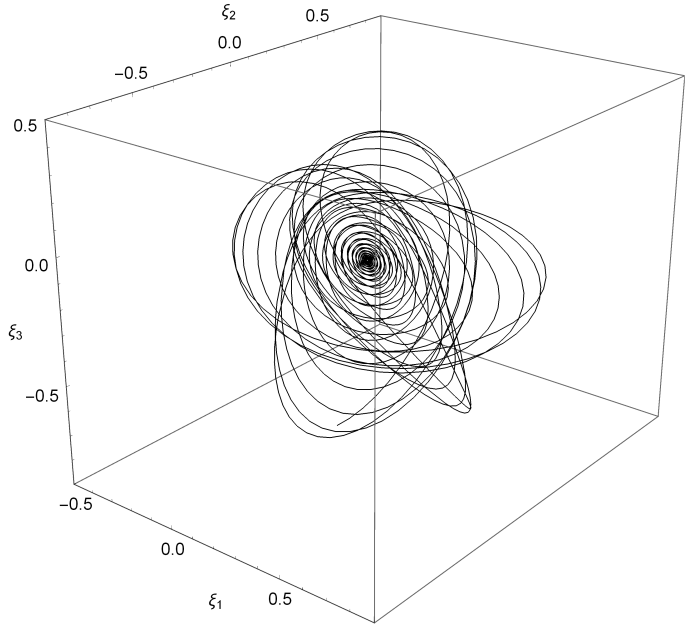}
\caption{The trajectory of the system (2.9) with ${\bm a}=(1,0,-1,0,2,-1,1,-1)$, 
${\bm b}=(0.1,0.1,0.1,0.1,0.1,0.1,0.1,0.1)$ and
${\bm\xi}_0=(0,0,\frac{\sqrt{3}}{2},0,0,0,0,\frac{1}{2})$ (pure state), of the dumped quasiperiodic
motion, going to the stationary (equilibrium) solution $\bar{\bm\xi}=(0.284966,-0.168841,-0.042086,-0.035279,
0.556160,-0.356250,0.522711,-0.421682)$ (pure state).}
\end{figure}
\section{Dynamics of entropy for qutrit states}
\label{sec:5}
Our purpose now is to study the temporal development of entropy under the 
evolution of the qutrit state given by (1.1), (2.7) and (2.8).  Solving the 
characteristic equation $\det[\rho(t)-\nu I]=0$, where $\rho(t)$ is given by
(2.7), that is cubic in $\nu$, we get the following trigonometric solution
\begin{equation}
\nu_1=\frac{1}{3}+\frac{2}{3}|{\bm\xi}(t)|\cos\frac{\alpha}{3},\quad 
\nu_{2,3}=\frac{1}{3}-\frac{2}{3}|{\bm\xi}(t)|\cos\left(\frac{\alpha}{3}\pm\frac{\pi}{3}\right),
\end{equation}
where
\begin{equation}
\cos\alpha=\frac{{\bm\xi}(t)\cdot[{\bm\xi}(t)*{\bm\xi}(t)]}{|{\bm\xi}(t)|^3}.
\end{equation}
Hence, we obtain the following expression for the entropy of the qutrit states
\begin{align}
S(t) =& -{\rm Tr}[\rho(t)\log_3\rho(t)]\nonumber\\
{} =& -\frac{1}{3}\left(1+2|{\bm\xi}(t)|\cos\frac{\alpha}{3}\right)\log_3\frac{1}{3} 
\left(1+2|{\bm\xi}(t)|\cos\frac{\alpha}{3}\right)\nonumber\\
&{}-\frac{1}{3}\left[1-2|{\bm\xi}(t)|\cos\left(\frac{\alpha}{3}+\frac{\pi}{3}\right)\right]
\log_3\frac{1}{3}\left[1-2|{\bm\xi}(t)|\cos\left(\frac{\alpha}{3}+\frac{\pi}{3}\right)\right]\nonumber\\
&{}-\frac{1}{3}\left[1-2|{\bm\xi}(t)|\cos\left(\frac{\alpha}{3}-\frac{\pi}{3}\right)\right]
\log_3\frac{1}{3}\left[1-2|{\bm\xi}(t)|\cos\left(\frac{\alpha}{3}-\frac{\pi}{3}\right)\right].
\end{align}
Notice that the maximum value 1 of the entropy $S(t)$ corresponds to 
$|{\bm\xi}(t)|=0$ i.e. maximally mixed state, and its minimum value 0 is reached at $|{\bm\xi}(t)|=1$, 
${\bm\xi}(t)*{\bm\xi}(t)={\bm\xi}(t)$ referring to the pure state (see (2.4)). The entropy is 
periodic function of time for periodic solutions to (2.9) corresponding to mixed 
states and as it follows from numerical calculations, decreasing function of time 
involving possibility of dumped oscillations, for solutions going to equilibrium 
points on the boundary of the space of states except of those with the initial data 
corresponding to pure states.  Further, the entropy is constant for stationary solutions 
and trajectories on the boundary $\partial\Omega_p$. The novel behavior of the entropy that is absent 
in the evolution of qubit states includes the case of the auto-oscillations connected 
with the existence of the limit cycles illustrated in Fig.\ 7.  Another example are the horizontal 
asymptotes corresponding to the asymptotically stable stationary solutions connected with 
the mixed states on the boundary of the state space.
\begin{figure*}
\centering
\includegraphics[width =.75\textwidth]{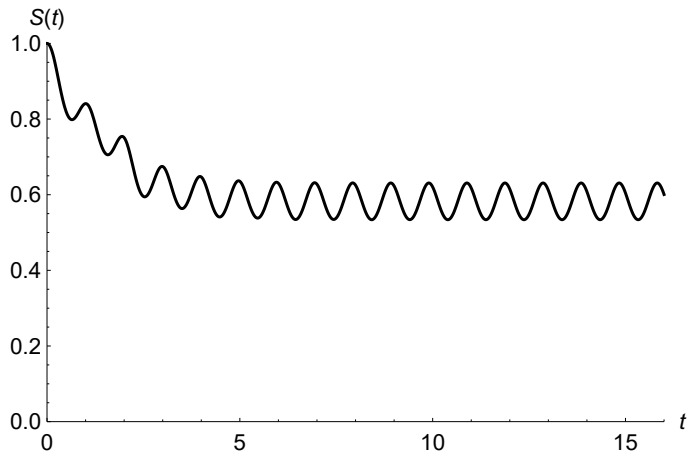}
\caption{Auto-oscillations of the entropy of the qutrit given by (5.3) connected with the existence of the 
limit cycle of the system (2.9), where the parameters are ${\bm a}=(1,1,0,2,-2,1,0,0)$, 
${\bm b}=(0,0,\frac{\sqrt{3}}{2},0,0,0,0,\frac{1}{2})$, and ${\bm\xi}_0=(0,0,0,0,0,0,0,0)$.}
\end{figure*}
\section{Conclusions}
In this work we study the nonlinear generalization of the von Neumann equation.  
An advantage of such nonlinear generalization in comparison with alternative 
approaches is that it preserves the convexity of the space of quantum states.  
This means that the introduced nonlinearity does not violate the probabilistic 
structure of quantum mechanics.  Another asset of the approach taken up in this work 
usually absent in nonlinear generalizations of quantum mechanics, is that 
it does not enable superluminal communication.  We point out that the nonlinear 
generalization of the von Neumann equation describes open quantum systems.  This 
fact was recognized in the special case of the qubit in Ref.\ 11 based on the 
applications of non-Hermitian Hamiltonians such as $H+{\rm i}G$ given by (1.10) 
in the theory of open quantum systems.  The authors presented therein as an illustrative 
example the observations of the classical paper of Feshbach \cite{18} who utilized for the 
first time non-Hermitian Hamiltonians for description of coupling between 
a discrete level and a continuum of states of a given quantum system.  Such 
approach is nowadays employed in physics of open quantum systems as for instance
phase transitions.  Another example is non-Hermiticity appearing in 
nononequilibrium open systems with gain and/or loss \cite{19,20,21,22,23}.
Even more convincing evidence of the openness of the quantum system described by 
the nonlinear von Neumann equation (1.2) is the global asymptotic stability of 
stationary solutions to the nonlinear system (1.12) mentioned in the 
Introduction, corresponding to evolution of the qubit from mixed states to the 
pure ones.  In other words, in the discussed case every mixed state from the interior 
of the Bloch sphere goes to the pure state represented by the point on the sphere.  
Of course, as depicted in Ref.\ 1 in Fig.\ 4, entropy is then decreasing function 
of time.  Clearly such situation cannot occur in the case of an isolated system.

As a concrete realization of the nonlinear von Neumann equation we have chosen the 
nontrivial case of the qutrit.  As one would expect, this three-level system exhibits 
richer dynamics than does the qubit analyzed in the recent work \cite{1}.  In particular, 
the classical Riccati system (2.9) corresponding to the qutrit has quasiperiodic solutions 
and limit cycles that are absent in the case of the qubit.  Interestingly, the 
quasiperiodic trajectories in the investigated three-level quantum system such 
as that illustrated in Fig.\ 3 have counterpart in classical orbital mechanics 
called Lissajous orbits that an object can follow around the Lagrangian point of a 
three-body system.  The existence of the limit cycles of the system (2.9) and 
consequently self-oscillations of the qutrit is remarkable. The author does not 
know any other example of the limit cycle in a simple quantum system like that 
described in this work.  For instance, one can find in the literature limit 
cycles in renormalization group behaviour of quantum Hamiltonians \cite{24}. 
Self-oscillations as dynamical peculiarity of open systems are well-known
in classical physics.  For example we can read in Ref.\ 25 that: ``Self-oscillation
system is an open system because emergence and maintenance energy is given by an
external source.'' or \cite{26}: ``Self-oscillatory systems have some properties 
that differ from those of harmonic oscillators.  First, they exhibit undamped 
oscillations by taking and dissipating energy from various sources; thus, the 
systems have typical characteristics of nonequilibrium open systems.''  Referring 
to the harmonic oscillator mentioned in the second quotation we point out that 
the closed, periodic trajectory depicted in Fig.\ 5 is reached only 
asymptotically. An example of connection between self-oscillations of energy and 
entropy of a mechanical system was provided in Ref.\ 27.  An important example of 
self-oscillations in open systems far from equilibrium is the temporal 
celebrated dissipative structures in the form of sustained oscillations, 
illustrated by biological rhythms \cite{28} playing the fundamental role in the 
biological self-organization.  Bearing in mind the existence of the limit cycle 
in the case of the qutrit and its lack for the qubit's time evolution, it is 
worthwhile to note that occurrence of dissipative structure in the case of 
chemical reactions takes place for three-particle reactions \cite{29} that are 
less probable than two-particle ones.  Referring back to the ringing structure
in the entropy in Fig.\ 7 we point out an intriguing interpretation of such entropy
oscillations \cite{30} as a process of self-organization and process of disorganization
of an open system arising after reaching some critical level of organization. It is further 
suggested that these oscillations can model action of mankind such as constructive one like:
``building houses and factories, partition off rivers by dams and etc.'' and destructive:
such as ``disintegration of ecosystems, destructive fluctuations of the climate.''
and so on.  As an advantage of this approach the author of Ref.\ 30 indicates that    
``The understanding of the reasons of the given tendency not only enables to foresee future
hardships, but also prompts a way to avoid them.''

The fundamental character of the notion of the qutrit in quantum information theory suggests 
that the observations of this work concerning the nonlinear evolution of the qutrit would be of 
importance in quantum information processing.
We also remark that some of the identities presented in Appendix concerning 
star and wedge products related to the structure constants of the $su(3)$ algebra are most 
probably new.  It seems that they would be a useful tool in the study of the 
qutrit and other problems connected with the $SU(3)$ symmetry.
\begin{acknowledgements}
I would like to thank Jakub Rembieli\'nski for helpful comments.
\end{acknowledgements}
\appendix
\section*{Appendix}
\setcounter{section}{1}
\setcounter{equation}{0}
We begin by collecting the basic properties of the Gell-Mann matrices
\begin{alignat*}{5}
\lambda_1=&
\begin{pmatrix}
0 & 1 & 0\\
1 & 0 & 0\\
0 & 0 & 0
\end{pmatrix},\quad &
\lambda_2=&
\begin{pmatrix}
0 & -{\rm i} & 0\\
{\rm i} & 0 & 0\\
0 & 0 & 0
\end{pmatrix},\quad &
\lambda_3=&
\begin{pmatrix}
1 & 0 & 0\\
0 & -1 & 0\\
0 & 0 & 0
\end{pmatrix},\quad &
\lambda_4=&
\begin{pmatrix}
0 & 0 & 1\\
0 & 0 & 0\\
1 & 0 & 0
\end{pmatrix},\quad &
\lambda_5=& 
\begin{pmatrix}
0 & 0 & -{\rm i}\\
0 & 0 & 0\\
{\rm i} & 0 & 0
\end{pmatrix},
\end{alignat*}
\begin{alignat}{3}
\lambda_6=&
\begin{pmatrix}
0 & 0 & 0\\
0 & 0 & 1\\
0 & 1 & 0
\end{pmatrix},\quad &
\lambda_7=&
\begin{pmatrix}
0 & 0 & 0\\
0 & 0 & -{\rm i}\\
0 & {\rm i} & 0
\end{pmatrix},\quad &
\lambda_8=&\frac{1}{\sqrt{3}}
\begin{pmatrix}
1 & 0 & 0\\
0 & 1 & 0\\
0 & 0 & -2
\end{pmatrix}.
\end{alignat}
These matrices are Hermitian and traceless.  They obey the following commutation 
relations
\begin{align}
[\lambda_j,\lambda_k]&=2{\rm i}f_{jkl}\lambda_l,\\
\{\lambda_j,\lambda_k\}&=\frac{4}{3}\delta_{jk}+2d_{jkl}\lambda_l,
\end{align}
where the coefficients (structure constants of the $su(3)$ algebra) $f_{jkl}$ are 
completely anti-symmetric and $d_{jkl}$ are completely symmetric.  The 
independent, nonvanishing components of $f_{jkl}$ and $d_{jkl}$ are given by
\begin{equation}
\begin{split}
f_{123}&=1,\,\,f_{458}=f_{678}=\frac{\sqrt{3}}{2},\\
f_{147}&=f_{246}=f_{257}=f_{345}=f_{516}=f_{637}=\frac{1}{2},
\end{split}
\end{equation}
and
\begin{align}
d_{118} &= d_{228}=d_{338}=-d_{888}=\frac{1}{\sqrt{3}},\nonumber\\
d_{448} &= d_{558}=d_{668}=d_{778}=-\frac{1}{2\sqrt{3}},\\
d_{146} &= d_{157}=-d_{247}=d_{256}=d_{344}=d_{355}=-d_{366}=-d_{377}=\frac{1}{2}\nonumber.
\end{align}
We define with the help of the coefficients $f_{ijk}$ and $d_{ijk}$, 
respectively, the anti-symmetric wedge product of vectors ${\bm a},\,{\bm b}\in{\mathbb R}^8$:
\begin{equation}
({\bm a}\wedge{\bm b})_i=\sqrt{3}f_{ijk}a_jb_k,
\end{equation}
and the symmetric star-product
\begin{equation}
({\bm a}*{\bm b})_i=\sqrt{3}d_{ijk}a_jb_k.
\end{equation}
Using (A.4) and (A.6) we find that the coordinates of the vector ${\bm a}\wedge{\bm b}$ are
\begin{equation}
\begin{split}
({\bm a}\wedge{\bm b})_1&=\sqrt{3}[a_2b_3-a_3b_2+\textstyle{\frac{1}{2}}(a_4b_7-a_7b_4)+
\textstyle{\frac{1}{2}}(a_6b_5-a_5b_6)],\\
({\bm a}\wedge{\bm b})_2&=\sqrt{3}[\textstyle{\frac{1}{2}}(a_4b_6-a_6b_4)+\textstyle{\frac{1}{2}}(a_5b_7-a_7b_5)+
a_3b_1-a_1b_3)],\\
({\bm a}\wedge{\bm b})_3&=\sqrt{3}[\textstyle{\frac{1}{2}}(a_4b_5-a_5b_4)+\textstyle{\frac{1}{2}}(a_7b_6-a_6b_7)+
a_1b_2-a_2b_1)],\\
({\bm a}\wedge{\bm b})_4&=\sqrt{3}[\textstyle{\frac{\sqrt{3}}{2}}(a_5b_8-a_8b_5)+\textstyle{\frac{1}{2}}(a_7b_1-a_1b_7)+
\textstyle{\frac{1}{2}}(a_6b_2-a_2b_6)+\textstyle{\frac{1}{2}}(a_5b_3-a_3b_5)],\\
({\bm a}\wedge{\bm b})_5&=\sqrt{3}[\textstyle{\frac{1}{2}}(a_1b_6-a_6b_1)+\textstyle{\frac{1}{2}}(a_3b_4-a_4b_3)+
\textstyle{\frac{1}{2}}(a_7b_2-a_2b_7)+\textstyle{\frac{\sqrt{3}}{2}}(a_8b_4-a_4b_8)],\\
({\bm a}\wedge{\bm b})_6&=\sqrt{3}[\textstyle{\frac{1}{2}}(a_3b_7-a_7b_3)+\textstyle{\frac{1}{2}}(a_5b_1-a_1b_5)+
\textstyle{\frac{1}{2}}(a_2b_4-a_4b_2)+\textstyle{\frac{\sqrt{3}}{2}}(a_7b_8-a_8b_7)],\\
({\bm a}\wedge{\bm b})_7&=\sqrt{3}[\textstyle{\frac{\sqrt{3}}{2}}(a_8b_6-a_6b_8)+\textstyle{\frac{1}{2}}(a_1b_4-a_4b_1)+
\textstyle{\frac{1}{2}}(a_2b_5-a_5b_2)+\textstyle{\frac{1}{2}}(a_6b_3-a_3b_6)],\\
({\bm a}\wedge{\bm b})_8&=\sqrt{3}[\textstyle{\frac{\sqrt{3}}{2}}(a_4b_5-a_5b_4)+\textstyle{\frac{\sqrt{3}}{2}}(a_6b_7-a_7b_6)].
\end{split}
\end{equation}
Furthermore, making use of (A.5) and (A.7) we get the coordinates of the vector ${\bm a}*{\bm b}$
\begin{equation}
\begin{split}
({\bm a}*{\bm b})_1&=\sqrt{3}[\textstyle{\frac{1}{\sqrt{3}}}(a_1b_8+a_8b_1)+\textstyle{\frac{1}{2}}(a_4b_6+a_6b_4)
+\textstyle{\frac{1}{2}}(a_5b_7+a_7b_5)],\\
({\bm a}*{\bm b})_2&=\sqrt{3}[\textstyle{\frac{1}{\sqrt{3}}}(a_2b_8+a_8b_2)-\textstyle{\frac{1}{2}}(a_4b_7+a_7b_4)
+\textstyle{\frac{1}{2}}(a_5b_6+a_6b_5)],\\
({\bm a}*{\bm b})_3&=\sqrt{3}[\textstyle{\frac{1}{\sqrt{3}}}(a_3b_8+a_8b_3)+\textstyle{\frac{1}{2}}(a_4b_4+a_5b_5
-a_6b_6-a_7b_7)],\\
({\bm a}*{\bm b})_4&=\sqrt{3}[-\textstyle{\frac{1}{2\sqrt{3}}}(a_4b_8+a_8b_4)+\textstyle{\frac{1}{2}}(a_1b_6+a_6b_1)
-\textstyle{\frac{1}{2}}(a_7b_2+a_2b_7)+\textstyle{\frac{1}{2}}(a_3b_4+a_4b_3)],\\
({\bm a}*{\bm b})_5&=\sqrt{3}[-\textstyle{\frac{1}{2\sqrt{3}}}(a_5b_8+a_8b_5)+\textstyle{\frac{1}{2}}(a_1b_7+a_7b_1)
+\textstyle{\frac{1}{2}}(a_2b_6+a_6b_2)+\textstyle{\frac{1}{2}}(a_3b_5+a_5b_3)],\\
({\bm a}*{\bm b})_6&=\sqrt{3}[-\textstyle{\frac{1}{2\sqrt{3}}}(a_6b_8+a_8b_6)+\textstyle{\frac{1}{2}}(a_1b_4+a_4b_1)
+\textstyle{\frac{1}{2}}(a_2b_5+a_5b_2)-\textstyle{\frac{1}{2}}(a_3b_6+a_6b_3)],\\
({\bm a}*{\bm b})_7&=\sqrt{3}[-\textstyle{\frac{1}{2\sqrt{3}}}(a_7b_8+a_8b_7)+\textstyle{\frac{1}{2}}(a_1b_5+a_5b_1)
-\textstyle{\frac{1}{2}}(a_2b_4+a_4b_2)-\textstyle{\frac{1}{2}}(a_3b_7+a_7b_3)],\\
({\bm a}*{\bm b})_8&=\sqrt{3}[\textstyle{\frac{1}{\sqrt{3}}}(a_1b_1+a_2b_2+a_3b_3-a_8b_8)
-\textstyle{\frac{1}{2\sqrt{3}}}(a_4b_4+a_5b_5+a_6b_6+a_7b_7)].
\end{split}
\end{equation}
In the particular case of ${\bm a}=(0,0,a_3,0,0,0,0,a_8)$ and ${\bm b}=(0,0,b_3,0,0,0,0,b_8)$ we have
\begin{equation}
\begin{split}
({\bm a}*{\bm b})_3=a_3b_8+a_8b_3,\\
({\bm a}*{\bm b})_8=a_3b_3-a_8b_8,
\end{split}
\end{equation}
where the remaining coordinates of ${\bm a}*{\bm b}$ vanish.
We now write down some identities satisfied by the wedge and the star products.  We 
begin with the relation implied by the Jacobi identity satisfied by the 
coefficients $f_{ijk}$ such that
\begin{equation}
{\bm a}\wedge({\bm b}\wedge{\bm c})+{\bm b}\wedge({\bm c}\wedge{\bm a})+{\bm c}\wedge({\bm a}\wedge{\bm b})=0.
\end{equation}
Using the identity \cite{14}
\begin{equation}
f_{ijk}f_{klm}=\frac{2}{3}(\delta_{il}\delta_{jm}-\delta_{im}\delta_{jl})+d_{ilk}d_{jmk}-d_{jlk}d_{imk},
\end{equation}
we get
\begin{equation}
{\bm a}\wedge({\bm b}\wedge{\bm c})=2[{\bm b}({\bm a}\cdot{\bm c})-{\bm c}({\bm a}\cdot{\bm b})]
+{\bm b}*({\bm a}*{\bm c})-{\bm c}*({\bm a}*{\bm b}),
\end{equation}
and
\begin{equation}
({\bm a}\wedge{\bm b})^2=\frac{2}{3}[{\bm a}^2{\bm b}^2-({\bm a}\cdot{\bm b})^2]
+\frac{1}{3}[({\bm a}*{\bm a})\cdot({\bm b}*{\bm b})-({\bm a}*{\bm b})^2].
\end{equation}
The next identity \cite{31}
\begin{equation}
f_{ijk}d_{klm}+f_{ilk}d_{jmk}+f_{imk}d_{jlk}=0,
\end{equation}
implies
\begin{equation}
{\bm a}\wedge({\bm b}*{\bm c})+{\bm b}\wedge({\bm c}*{\bm a})+{\bm c}\wedge({\bm a}*{\bm b})=0.
\end{equation}
Now, taking into account the identity \cite{14}
\begin{equation}
f_{ijk}d_{klm}+f_{ljk}d_{imk}+f_{mjk}d_{ilk}=0,
\end{equation}
we find
\begin{equation}
{\bm a}\wedge({\bm b}*{\bm c})+{\bm c}*({\bm b}\wedge{\bm a})+{\bm b}*({\bm c}\wedge{\bm a})=0.
\end{equation}
An immediate consequence of (A.16) or (A.18) is
\begin{equation}
{\bm a}\wedge({\bm a}*{\bm a})=0.
\end{equation}
Taking into account (A.19) one can show \cite{16} that
\begin{equation}
{\bm a}\wedge{\bm b}=0\iff {\bm b}=\mu{\bm a}+\nu{\bm a}*{\bm a},
\end{equation}
where $\mu$ and $\nu$ are constant.  Further, (A.16) and (A.18) taken together yield
\begin{equation}
{\bm b}\wedge({\bm a}*{\bm c})+{\bm c}\wedge({\bm a}*{\bm b})
={\bm c}*({\bm b}\wedge{\bm a})+{\bm b}*({\bm c}\wedge{\bm a}).
\end{equation}
Finally, the identity \cite{31}
\begin{equation}
d_{ijk}d_{klm}+d_{ilk}d_{kjm}+d_{imk}d_{kjl}=\frac{1}{3}(\delta_{ij}\delta_{lm}+\delta_{il}\delta_{jm}
+\delta_{im}\delta_{jl}),
\end{equation}
leads to
\begin{equation}
{\bm a}*({\bm b}*{\bm c})+{\bm b}*({\bm c}*{\bm a})+{\bm c}*({\bm a}*{\bm b})
={\bm a}({\bm b}\cdot{\bm c})+{\bm b}({\bm c}\cdot{\bm a})+{\bm c}({\bm a}\cdot{\bm b}).
\end{equation}
In the special case ${\bm a}={\bm b}={\bm c}$ we get from (A.23) the useful 
relation
\begin{equation}
{\bm a}*({\bm a}*{\bm a})={\bm a}^2{\bm a}.
\end{equation}
On the other hand, using (A.23) and (A.24) we obtain
\begin{equation}
({\bm a}*{\bm a})*({\bm a}*{\bm a})=2[{\bm a}\cdot({\bm a}*{\bm a})]{\bm a}-{\bm a}^2{\bm a}*{\bm a}.
\end{equation}
We supplement the identities concerning wedge and star product with the following 
quite obvious relations involving scalar product
\begin{align}
{\bm a}\cdot({\bm b}\wedge{\bm c})&={\bm b}\cdot({\bm c}\wedge{\bm a})={\bm c}\cdot({\bm a}\wedge{\bm b}),\\
{\bm a}\cdot({\bm b}*{\bm c})&={\bm b}\cdot({\bm c}*{\bm a})={\bm c}\cdot({\bm a}*{\bm b}).
\end{align}
From (A.24) and (A.27) it follows that
\begin{equation}
({\bm a}*{\bm a})^2=|{\bm a}|^4,
\end{equation}
where $|{\bm a}|=\sqrt{{\bm a}^2}$ is the Euclidean norm of the vector ${\bm a}$.

We end this section with algebraic relations for the Gell-Mann matrices.  We 
first write down the formula for the product of these matrices following directly 
from (A.2) and (A.3)
\begin{equation}
\lambda_j\lambda_k=\frac{2}{3}\delta_{jk}+(d_{jkl}+{\rm i}f_{jkl})\lambda_l.
\end{equation}
Hence, we get
\begin{equation}
({\bm a}\cdot{\bm\lambda})({\bm b}\cdot{\bm\lambda})=\frac{2}{3}{\bm a}\cdot{\bm b}
+\left[\frac{1}{\sqrt{3}}{\bm a}*{\bm b}+\frac{{\rm i}}{\sqrt{3}}{\bm a}
\wedge{\bm b}\right]\cdot{\bm\lambda}.
\end{equation}
As an immediate consequence of the relation (A.30), we have
\begin{align}
&({\bm a}\cdot{\bm\lambda})^2=\frac{2}{3}{\bm a}^2 + \frac{1}{\sqrt{3}}({\bm 
a}*{\bm a})\cdot{\bm\lambda},\\
&[{\bm a}\cdot{\bm\lambda},{\bm b}\cdot{\bm\lambda}]=\frac{2{\rm i}}{\sqrt{3}}
({\bm a}\wedge{\bm b})\cdot{\bm\lambda},\\
&\{{\bm a}\cdot{\bm\lambda},{\bm b}\cdot{\bm\lambda}\}=\frac{4}{3}{\bm a}\cdot{\bm b}
+\frac{2}{\sqrt{3}}({\bm a}*{\bm b})\cdot{\bm\lambda}.
\end{align}
Using (A.13) and (A.30) we obtain
\begin{align}
&({\bm a}\cdot{\bm\lambda})({\bm b}\cdot{\bm\lambda})({\bm c}\cdot{\bm\lambda})=
\frac{2}{3\sqrt{3}}[{\bm c}\cdot({\bm a}*{\bm b})+{\rm i}{\bm c}\cdot({\bm a}
\wedge{\bm b})]\nonumber\\
&{}+\left\{-\frac{1}{3}[{\bm b}*({\bm a}*{\bm c})-{\bm c}*({\bm a}*{\bm b})
-{\bm a}*({\bm b}*{\bm c})]
-\frac{2}{3}[{\bm b}({\bm a}\cdot{\bm c})-{\bm c}({\bm a}\cdot{\bm b})
-{\bm a}({\bm b}\cdot{\bm c})]\right.\nonumber\\
&\quad\left.{}+\frac{\rm i}{3}[{\bm c}*({\bm a}\wedge{\bm b})-{\bm c}\wedge({\bm a}*{\bm b})]
\right\}\cdot{\bm\lambda}.
\end{align}
For ${\bm c}={\bm a}$ the formula (A.34) reduces to
\begin{equation}
({\bm a}\cdot{\bm\lambda})({\bm b}\cdot{\bm\lambda})({\bm a}\cdot{\bm\lambda})=
\frac{2}{3\sqrt{3}}{\bm a}\cdot({\bm a}*{\bm b})+\left[-\frac{2}{3}{\bm b}*({\bm a}*{\bm a})
-\frac{1}{3}{\bm a}^2{\bm b}+2({\bm a}\cdot{\bm b}){\bm a}\right]\cdot{\bm\lambda}
\end{equation}
following directly from (A.18) and (A.23).  A direct consequence of (A.24) and (A.35) is
the relation 
\begin{equation}
({\bm a}\cdot{\bm\lambda})^3=\frac{2}{3\sqrt{3}}{\bm a}\cdot({\bm a}*{\bm a})+{\bm a}^2{\bm a}
\cdot{\bm\lambda}.
\end{equation}
We finally write down the trace-orthogonality relations
\begin{equation}
{\rm Tr}(\lambda_j\lambda_k)=2\delta_{jk},
\end{equation}
and the invariants of the $SU(3)$ group related to the Casimir operators such 
that
\begin{align}
\frac{1}{2}{\rm Tr}({\bm a}\cdot{\bm\lambda})^2 &={\bm a}^2,\\
\frac{\sqrt{3}}{2}{\rm Tr}({\bm a}\cdot{\bm\lambda})^3 &={\bm a}\cdot({\bm a}*{\bm a})
\end{align}
implied by (A.31), (A.36), and (A.37).  The second invariant (A.39) is related to 
the determinant of the matrix ${\bm a}\cdot{\bm\lambda}$ by
\begin{equation}
\det({\bm a}\cdot{\bm\lambda})=\frac{2}{3\sqrt{3}}{\bm a}\cdot({\bm a}*{\bm a}).
\end{equation}
This invariant can be written in the coordinate form as
\begin{align}
&{\bm a}\cdot({\bm a}*{\bm a})=3a_8(a_1^2+a_2^2+a_3^2)-a_8^3-\frac{3}{2}a_8(a_4^2+a_5^2
+a_6^2+a_7^2)\nonumber\\
&{}+\frac{3\sqrt{3}}{2}a_3(a_4^2+a_5^2-a_6^2-a_7^2)+3\sqrt{3}[(a_1a_6-a_2a_7)a_4+(a_1a_7+a_2a_6)a_5].
\end{align}


%
%


\begin{thebibliography}{}
\bibitem{1}Kowalski, K., Rembieli\'nski, J.: Integrable nonlinear evolution of the qubit. Ann. Phys. {\bf 411}, 167955 (2019)
\bibitem{2}Gisin, N.: Irreversible quantum dynamics and the Hilbert space structure of quantum kinematics. J. Math. Phys. 
{\bf 24}(7), 1779 (1983)
\bibitem{3}Turski, {\L}.A.: Dissipative quantum mechanics. Metriplectic dynamics in action. In: From Quantum Mechanics to 
Technology. Lecture Notes in Physics {\bf 477}, 347 (2007)
\bibitem{4}Grigorenko, A.N.: Measurement description by means of a nonlinear Schrodinger equation. J. Phys. A: Math. Gen. 
{\bf 28}, 1459 (1995) 
\bibitem{5}Kraus, K.: States, Effects, and Operations: Fundamental Notions of Quantum Theory, Springer, Berlin (1983)
\bibitem{6}Grabowski, J., Ku{\'s}, M., Marmo, G.: Symmetries, group actions, and entanglement. Open Sys. \& Information Dyn. 
{\bf 13}, 343 (2006)
\bibitem{7}Rembieli\'nski, J., Caban, P.: Nonlinear evolution and signaling. Phys. Rev. Research {\bf 2}, 012027(R) (2020)
\bibitem{8}Di{\'o}si, L.: Nonlinear Schr\"odinger equation in foundations: summary of 4 catches. J. Phys. Conf. Ser. 
{\bf 701}, 012019 (2016)
\bibitem{9}Lamb, W.E., Jr.: Theory of an Optical Maser. Phys. Rev. {\bf 134}, A1429 (1964)
\bibitem{10}Bargmann, V., Michel, L., Telegdi, V.L.: Precession of the polarization of particles moving in a homogeneous 
electromagnetic field. Phys. Rev. Lett. {\bf 2}, 435 (1959) 
\bibitem{11}Grimaudo, R., de Castro, A.S.M., Ku\'s, Messina, A.: Exactly solvable time-dependent pseudo-Hermitian su(1,1) 
Hamiltonian models. Phys. Rev. A {\bf 98}, 033835 (2018)
\bibitem{12}Rembieli\'nski, J., Caban, P.: Nonlinear extension of the quantum dynamical semigroup, 2020 (unpublished)
\bibitem{13}Goyal, S.K., Simon, B.N.,  Singh, R., Simon, S.: Geometry of the generalized Bloch 
sphere for qutrits. J. Phys. A: Math. Theor. {\bf 49}, 165203 (2016)
\bibitem{14}Arvind, Mallesh, K.S., Mukunda, N.: A generalized Pancharatnam geometric phase formula for 
three-level quantum systems. J. Phys. A: Math. Gen. {\bf 30}, 2417 (1997)  
\bibitem{15}Korn, G.A., Korn, T.M.: Mathematical Handbook for Scientists and Engineers, Dover, New York (2000)
\bibitem{16}Mallesh, K.S., Mukunda, N.: The algebra and geometry of $SU(3)$ matrices. Pramana {\bf 49}, 371 (1997) 
\bibitem{17}B\"ol\"ukbal, A., Dereli, T.: On the $SU(3)$ parametrization of qutrits. J. Phys. Conf. Ser. {\bf 36}, 
28 (2006)
\bibitem{18}Feshbach, H.: Unified theory of nuclear reactions. Ann. Phys. {\bf 5}, 375 (1958); {\bf 19}, 287 (1962)
\bibitem{19}Konotop, V.V., Yang, J., Zezyulin, D.A.: Nonlinear Waves in PT -Symmetric Systems. Rev. Mod. Phys. {\bf 88},
035002 (2016)
\bibitem{20}Feng, L., El-Ganainy, R., Ge, L.: Non-Hermitian Photonics Based on Parity-Time Symmetry. Nat. Photon.
{\bf 11}, 752 (2017)
\bibitem{21}El-Ganainy, R., Makris, K.G., Khajavikhan, M., Musslimani, Z.H., Rotter, S., Christodoulides, D.N.: 
Non-Hermitian Physics and PT Symmetry. Nat. Phys. {\bf 14}, 11 (2018)
\bibitem{22}Miri, M.-A., Al{\'u}, A.: Exceptional Points in Optics and Photonics. Science {\bf 363}, 7709 (2019)
\bibitem{23}\"Ozdemir, \c{S}.K., Rotter, S., Nori, F., Yang, L.: Parity-Time Symmetry and Exceptional Points 
in Photonics. Nat. Mater. {\bf 18}, 783 (2019)
\bibitem{24}G{\l}azek, S.D.: Limit Cycles in Quantum Mechanics. In: Mathematical Physics of Quantum Mechanics. 
Lecture Notes in Physics {\bf 690}, 65 (2006)
\bibitem{25}Alkhasova, D.A., Sokotushchenko, V.N., Torchinsky, V.M., Zaichenko, V.M.: Peculiarities of Excitation of 
Self-Oscillations in Geological Systems. IOP Conf. Ser.: Earth Environ. Sci. {\bf 249}, 012026 (2019)
\bibitem{26}Kinoshita, S.: Pattern Formations and Oscillatory Phenomena, Elsevier, Amsterdam (2013)
\bibitem{27}Petrov, V.V., Ageev, V.M.: Entropy and auto-oscillations in nonlinear systems. Dokl. Akad. Nauk
SSSR {\bf 199}, 1260 (1971)
\bibitem{28}Goldbeter, A.: Dissipative structures in biological systems: bistability, oscillations, spatial patterns
and waves. Phil. Trans. R. Soc. A {\bf 376}, 20170376 (2018)
\bibitem{29}Nicolis, G., Prigogine, I., Self-organization in Nonequilibrium Systems. Wiley, New York (1977)
\bibitem{30}Shapovalov, V.I.: The criterion of ordering and self-organization of open system. Entropy oscillations 
in linear and nonlinear processes. International Journal of Applied Mathematics 
and Statistics {\bf 26}, 16-29 (2012); Shapovalov, V.I.: Entropy oscillations, ArXiv:0812.4031 
\bibitem{31}Haber, H.E.: Useful relations among the generators in the defining and adjoint representations of $SU(N)$ (2017)
(unpublished) http://scipp.ucsc.edu/~haber/ph218/sunid17.pdf  
\end{thebibliography}

%
%
\end{document}